%% file: ms.tex
\documentclass[preprint]{aastex701}

\usepackage{amssymb}
\usepackage{amsmath}
\usepackage{color}
\usepackage{xspace} 
\usepackage{ulem}
\usepackage{natbib}

\usepackage{macro_kh}

\newcommand\etacar{$\eta$ Carinae\xspace}

\newcommand\XRISM{{\it XRISM}\xspace}
\newcommand\rsl{{\it Resolve}\xspace}
\newcommand\xtd{{\it Xtend}\xspace}

\newcommand\EW{{\it EW}\xspace}

\shorttitle{XRISM observations of \etacar}
\shortauthors{XRISM Collaboration et al.}

\begin{document}

\title{High Spectral Resolution X-ray Observations of the Evolved Supermassive Stellar Binary System \etacar\ --- Iron K$\alpha$ Band Profile Revealed with \XRISM}

\correspondingauthor{Kenji Hamaguchi}
\email{kenji.hamaguchi@umbc.edu}

\collaboration{0}{XRISM Collaboration:}

\input{authorlist}

\begin{abstract}
The supermassive binary system, \etacar,
is experiencing enormous wind-driven mass loss 
at a rate unparalleled in the rest of the Galaxy.
Their wind-wind collision (WWC)
continuously produces shock heated, X-ray emitting plasmas.
The \XRISM\ X-ray observatory observed the system in 2023 and 2024 when the X-ray emission began to increase toward periastron passage in 2025. 
This manuscript reports unprecedentedly high-resolution X-ray spectra
in the iron K$\alpha$ band between 6.2 and 7.1~keV, 
obtained with the \rsl X-ray microcalorimeter.
The hydrogen-like (Ly$\alpha$) and helium-like (He$\alpha$) lines reveal three velocity components. 
Two of them are broadened with maximum velocities of 2000$-$3000~\UNITVEL,
likely originating from the post-shock companion wind.
The other is relatively narrow, with a Gaussian broadening of only  $\sim$290~\UNITVEL\ in 1 sigma, which may originate from the post-shock 
companion wind at the WWC stagnation point or penetrating the primary wind.
The iron fluorescent lines exhibit a moderate blueshift and broadening 
with velocities at 100$-$200~\UNITVEL, 
consistent with the primary wind's velocity field.
The spectra also confirm a Compton shoulder of the He$\alpha$ line complex for the first time.
Both fluorescing and scattering spectral profiles indicate that 
the binary system is seen from the companion side during these observations.
The flux ratio of the Compton scattering emission to the fluorescent line 
suggests substantial hydrogen depletion of the primary wind,
expected from CNO-cycled hydrogen nuclear fusion gas.
\end{abstract}


\section{Introduction}
\label{sec:intro}
Evolved massive stars lose a significant amount of mass through their winds \citep[e.g.,][]{Groh2014a}.
Their winds disperse nuclear-processed materials that emerge from 
the stellar interior to the surface, 
changing the circumstellar chemical composition. 
These processes mainly occur in the optical and infrared bands, 
but often evolve into high-energy phenomena as these stars tend to have 
a massive companion star, and their winds at a few thousand \UNITVEL\ 
collide with each other.
The wind-wind collision (WWC) produces a strong shock and thermalizes gas to an X-ray emitting temperature at tens of millions of degrees Kelvin.
Since X-ray observations provide the post-shock plasma conditions, and
UV, optical, or infrared observations can provide information on the unshocked wind conditions (e.g., terminal velocity and mass loss rate),
the WWC offers an ideal laboratory for studying astrophysical shock phenomena.
The WWC X-rays also irradiate the surrounding materials,
which then re-emit fluorescing or scattering X-rays.
This provides an independent mean of probing the unshocked winds.

Theoretical studies suggest that 
the WWC occurs in a conical region where the momenta of two winds balance,
and which wraps around the star with the weaker wind \citep[][]{Luo1990a,Usov1992b,Stevens1992}.
Each hypersonic wind produces a shock, 
which heats post-shock gas in proportion to the square of 
the wind velocity, \KT~$\sim$2~keV for $v$~=1000~\UNITVEL.
When the shock is adiabatic, that is, the post-shock flow time
is longer than the radiative cooling time,
the post-shock plasma flows along the WWC contact discontinuity
without significant cooling.
In the opposite case, when the shock is radiative,
the post-shock plasma collapses via quick cooling.
In the adiabatic limit, the luminosity should scale inversely with 
the binary separation.

The evolved supermassive star \etacar forms a binary system with
another unseen massive companion, 
driving the most vigorous WWC activity in the solar neighborhood
\citep[$d \sim$2.35~kpc,][]{Damineli1996,Ishibashi1999,Smith2006b}.
The primary star with an estimated mass at $\gtrsim$90~\UNITSOLARMASS\
currently flows an enhanced mass loss wind \citep[$v_{wind} \sim$420~\UNITVEL, \Mdot\ $\sim$8.5$\times$10$^{-4}$~\UNITSOLARMASSYEAR,][]{Davidson1995,Cox1995,Hillier2001,Groh2012}.
The star once experienced an extreme mass loss event called the Great Eruption 
in the 1840s, which ejected several tens of solar masses of nitrogen-rich materials, now seen as the bipolar Homunculus nebula \citep{Morris2017a,Smith2003b}.
The star is thus classified as an unstable Luminous Blue Variable 
({\it LBV}) phase.
The historical ejecta and the current thick wind heavily obscure
the companion.
However, X-rays and optical monitoring observations of the star 
since the late 1990s have found periodic variations from the WWC plasma,
revealing the presence of a massive companion star in a long, highly 
eccentric orbit 
\citep[$P$ =5.54~year, $e \sim$0.9,][]{Damineli2008,Corcoran2017a,Espinoza2022a}.
The observed X-ray characteristics (\KT~$\sim$4~keV, \LX~$\sim$10$^{35}$~\UNITLUMI) indicate that the companion star has 
a relatively thin, fast wind \citep[$v_{wind} \sim$3000~\UNITVEL, \Mdot\ $\sim$ 10$^{-5}$~\UNITSOLARMASSYEAR,][]{Pittard2002}.
These wind characteristics and high ionization UV and optical emission from reflection nebulae 
suggest that the companion is an O supergiant or nitrogen-rich Wolf-Rayet (WN) star \citep{VernerE2005a,Mehner2010b}.

Because of its importance for understanding the stellar evolution and shock physics, 
\etacar has been a frequent target of X-ray observatories 
\citep[][]{Corcoran1995,Tsuboi1997,CorcoranMF1998a,Corcoran1997,Ishibashi1999,Corcoran2000,Corcoran2001a,Leutenegger2003,Behar2007a,Hamaguchi2007b,Henley2008,Leyder2008,Sekiguchi2009,Leyder2010,Hamaguchi2014a,Hamaguchi2014b,Hamaguchi2016a,Corcoran2017a,Hamaguchi2018a,Kashi2021a,Espinoza2022a}.
Among them, the \CHANDRA\ Observatory's High-Energy Transmission Grating Spectrometer (HETGS) provided the highest resolution \etacar spectra,
which clearly separated emission lines from highly ionized Mg, Si, S, and Ca ions \citep{Corcoran2001a,Behar2007a,Henley2008,Espinoza2022a}.
The Si and S emission lines exhibited strong blue-shifted tails up to $\sim$2000~\UNITVEL,
which evolved as the binary system approached periastron \citep{Behar2007a}.
Their line centroids also vary with the orbit motion, which, however, cannot be reproduced with simple geometrical models,
in which the post-shock plasma flows constantly from the apex on a conical surface \citep{Henley2008}.

The Fe He$\alpha$ and Ly$\alpha$ emission lines between 6$-$7~keV trace \etacar's hottest WWC plasma near the apex,
which should provide crucial information on the WWC shock physics.
Unfortunately, \CHANDRA/HETGS has limited energy resolution ($\sim$30~eV) and collecting area ($\sim$30~cm$^{2}$) at 6~keV\footnote{https://cxc.harvard.edu/proposer/POG/html/chap8.html},
barely resolving the Fe He$\alpha$ line complex and the fluorescent 
K$\alpha$ line \citep[e.g.,][]{Corcoran2001a}.
\XRISM/\rsl features a superb energy resolution of $\sim$4.5~eV and a collecting area of $\sim$180~cm$^{2}$ at 6~keV \citep{Tashiro2025a}.
The \rsl\ spectra of \etacar\ reveal multiple velocity components in the iron emission lines as well as faint but important spectral features,
such as the Compton shoulders and absorption edges, in unprecedented detail.
This paper reports the results of the initial \XRISM observations of \etacar, conducted during the observatory's 
commissioning phase in 2023 and performance verification phase in 2024.

\section{Observations \& Data Reduction} \label{sec:obs}

The \XRISM X-ray observatory was launched from the Tanegashima Space Center in Japan on September 6, 2023 (UTC).
It has two X-ray telescopes, \rsl and \xtd, each comprising an X-ray Mirror Assembly (XMA) and a focal plane X-ray instrument \citep{Tashiro2025a}.
The XMAs of the two telescopes are nearly identical \citep{Hayashi2024a}.
Each XMA employs conically approximated Wolter I optics with 203 nested shells, focusing X-rays in a point-spread function (PSF) with a half-power diameter 
of 1.3$'-$1.4$'$.
{\rsl}'s focal plane instrument is an X-ray microcalorimeter array, comprising 35 pixels of 30\ARCSEC$\times$30\ARCSEC\ each 
with a total field of view (FOV) of 3.1\ARCMIN$\times$3.1\ARCMIN\ \citep{Kelley2024a}.
The detector is operated at 50~milli-Kelvin to detect a slight temperature increase by absorption of an X-ray photon.
This technology enables a superb energy resolution of $\sim$4.5~eV at 6~keV 
with High-resolution Primary (Hp) events, each of which does not have 
another event within 70.7~milliseconds.
\rsl is designed to be sensitive between 0.3$-$12~keV.
However, the detector gate valve used to keep a vacuum during launch had not successfully opened in orbit, limiting sensitivity during these observations to X-ray energies above 1.7~keV and reducing the collecting area to $\sim$60\% of the area with the gate valve opened at 6~keV \citep{Porter2024a}.
{\xtd}'s focal-plane instrument is an X-ray CCD camera with four identical CCD chips aligned squarely, covering a wide 38.5\ARCMIN$\times$38.5\ARCMIN\ field of view \citep{Mori2024a}.
The CCD chips employ backside-illuminated, p-channel technology, which enables high quantum efficiency between 0.4$-$13~keV.
The aim point is approximately 4.7\ARCMIN\ from the two inner chip edges on CCD\_ID =1, so that the main target is not placed on a chip gap.
The CCDs nominally take a frame exposure every 4~seconds. However, two CCDs,
CCD\_ID =0 \& 1 can continuously read only one-eighth of the field of view
every 0.5~second, in order to avoid photon pileup for bright sources and/or 
increase the time resolution.

\input{tab1}

\begin{figure}
\plotone{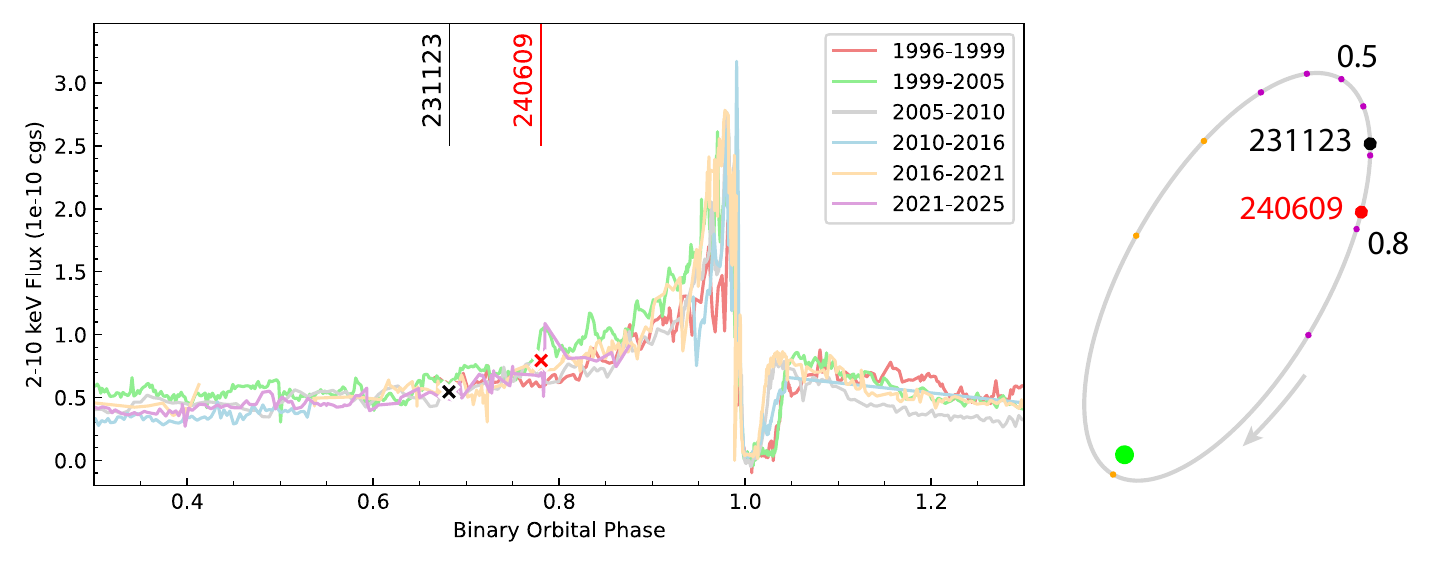}
\caption{{\it Left}: \XRISM \etacar\ observations in the 2$-$10~keV flux modulation plot folded with the binary orbital period,
obtained with \RXTE, \SWIFT, and \NIC\ since 1996 
\citep[][--- a complete analysis, including \NIC\ data obtained after 2021, 
will be presented in a forthcoming manuscript.]{Corcoran2017a,Espinoza2022a}
The crosses show the 2$-$10~keV fluxes during 
the \XRISM\ observations ({\it black}: XRI231123, {\it red}: XRI240609),
which are similar to those observed during earlier epochs near their orbital phases.
{\it Right}: Companion star's locations in a binary orbit during the \XRISM\ observations.
\label{fig:obs}
}
\end{figure}

\XRISM observed \etacar twice, starting on 2023 November 23 at the binary orbital phase $\phi_{\rm X} =$0.68 during the observatory's commissioning phase and
on 2024 June 9 at $\phi_{\rm X} =$0.78 during the performance verification phase, where the binary ephemeris is from \citet{Corcoran2017a} equation 4 
(the cycle count $E~=\phi_{\rm X}+4.0$ for these observations, 
see also Table~\ref{tbl:obslogs} and Figure~\ref{fig:obs}).
Hereafter, we abbreviate each observation with the last 2 digit of the year, two digits for the month, and the date, each after XRI, 
e.g., XRI231123 for the 2023 observation, XRI240609 for the 2024 observation.
XRI231123 was conducted for telescope pointing calibration without \rsl gain data acquisition during the observation (see Appendix~\ref{app:gaincal} for details).
The data without proper calibration leaves a $\sim$20~eV absolute gain offset and an additional instrumental broadening.
The \rsl instrument team made a special gain calibration of the data using $^{55}$Fe gain calibration obtained during the previous observation,
which improves the spectral gain accuracy to $\lesssim$1~eV.
XRI240609 was performed with an efficient gain calibration,
achieving a gain accuracy of $\lesssim$0.1~eV \citep{Porter2024a}.
\xtd\ used the one-eighth window option for XRI231123
and the full window option for XRI240609.
The latter dataset suffered from significant pileup near the PSF core of \etacar, and therefore, 
we use \xtd data for checking potential contamination 
by nearby X-ray sources.
XRI240606 \xtd observation detected a moderate X-ray flare from a young stellar object at 13.8\ARCMIN\ from \etacar\ \citep{Yoshida2024a}, which had no impact on the analysis of the \rsl\ spectrum.
Otherwise, no source in the Xtend FOV significantly contaminated the \rsl\ \etacar data.

We reprocessed the XRI240609 data with {\tt xapipeline} in HEASoft version 6.35.1, applying CALDB 20250315,
while we use the XRI231123 \rsl data reprocessed by the \XRISM data processing team with HEASoft version 6.34
and using the tailored gain history file (Appendix~\ref{app:gaincal}).
For both datasets, we followed the analysis instructions in the \XRISM data reduction 
guide\footnote{https://heasarc.gsfc.nasa.gov/docs/xrism/analysis/abc\_guide/xrism\_abc.html}
version~1.0.
We screened the data using standard criteria, selecting intervals above the geomagnetic cutoff rigidity indicator CORTIME above 4,
which encompasses all good time intervals that meet the standard criteria. 
For the \rsl spectral analysis, we selected Hp-grade events with the highest energy resolution
and exclude pixel-pixel coincident events.
We produced spectra from all active pixels but pixel 27, 
which occasionally shows unpredictable gain variations.
The Hp count rates of 34~pixels between 1.7$-$12~keV are 0.764~\UNITCPS\ during XRI231123 and 1.11~\UNITCPS\ during XRI240609.
At these low \rsl count rates, almost all Low-resolution Secondary (Ls) events are expected to originate from the particle background. 
We, therefore, excluded them in calculating the Hp event branching ratio for spectral response ({\tt rmf}) production.

We evaluated the non-X-ray background (NXB) contribution in the \rsl data using {\tt rslnxbgen} by following the procedure described in the ``NXB Spectral Models" 
page\footnote{https://heasarc.gsfc.nasa.gov/docs/xrism/analysis/nxb/nxb\_spectral\_models.html}.
We extracted an Hp-grade background spectrum from version~2 of the night Earth database\footnote{https://heasarc.gsfc.nasa.gov/docs/xrism/analysis/nxb/index.html}, obtained within $\pm$150~days of each observation,
outside of the South Atlantic Anomaly passage, with Earth elevation less than $-$5~\DEGREE,
Day Earth elevation more than 5~\DEGREE, and CORTIME above 4.
We then fit the spectrum to the v1 model ({\tt rsl\_nxb\_model\_v1.mo}) paired with the diagonal matrix, {\tt newdiag60000.rmf},
using the X-ray fitting package {\tt XSPEC} \citep{Arnaud1996}.
We added the best-fit NXB model to the source model for spectral fittings.
The NXB spectrum in the Fe K$\alpha$ band includes very weak Fe fluorescent lines;
however, the NXB contribution at any relevant energy range is less than 0.5\% of the source spectrum.
The cosmic-ray background (CXB) is $<$25\% of the NXB, so we ignored the CXB contribution.

\section{Results}\label{sec:res}

XRI240609 is the longest observation of \etacar with any sensitive X-ray focusing telescope.
Nevertheless, \etacar did not show any significant X-ray time variation during this observation.
We produced a 1.7$-$12~keV \rsl light curve of the observation with 200-second time bins, 
each with more than 80\% (160 sec) on-source time.
A constant model reproduces the light curve very well with ${\chi}^{2}$/dof = 1383.0/1450, 
where dof is the degrees of freedom.
XRI231123 similarly does not show significant variation with ${\chi}^{2}$/dof = 222.8/207.
These results are similar to earlier X-ray observations of \etacar, 
which do not show any short-term variation except near periastron \citep[e.g.,][]{Hamaguchi2007b,Hamaguchi2016a}. 

\begin{figure}
\plotone{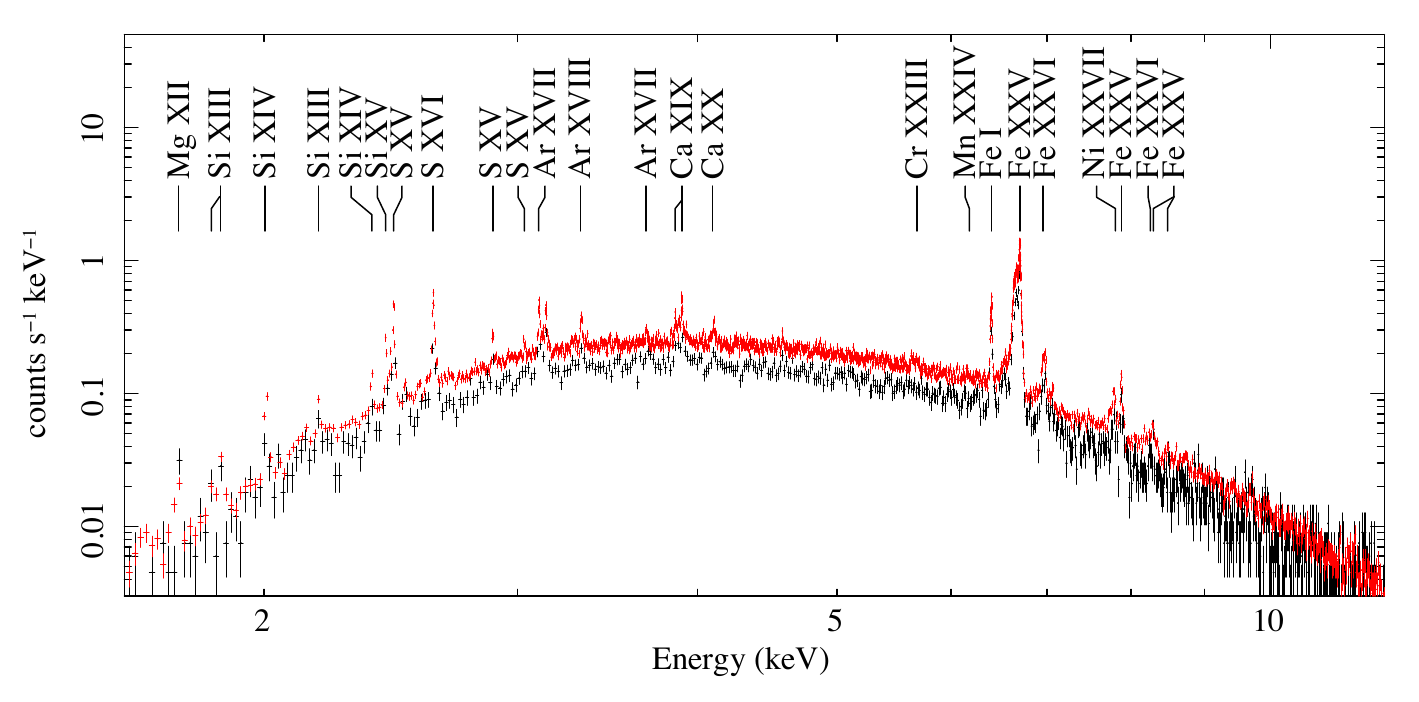}
\caption{\rsl spectra of \etacar during XRI231123 ({\it black}) and XRI240609 ({\it red}).
\label{fig:wholespec}
}
\end{figure}

Since \etacar does not exhibit significant variation in either observation,
we produced a spectrum from the entire dataset for each observation (Figure~\ref{fig:wholespec}).
The two spectra are very similar except for their fluxes: 
a simple model fit yields 2$-$10~keV fluxes of 5.5$\times$10$^{-11}$~\UNITFLUX\ for XRI231123
and 7.9$\times$10$^{-11}$~\UNITFLUX\ for XRI240609.
They exhibit strong line emission from highly ionized ions of multiple elements, Si, S, Ar, Ca, Fe and Ni.
They also show minor enhancements at some energies, which may be emission lines of rare elements, such as Cr and Mn.
These spectra are similar to earlier \CHANDRA\ high-resolution spectra \citep[e.g.,][]{Corcoran2001a}, but 
with significantly enhanced energy resolution and throughput.
The \rsl spectra separated the emission lines and display their profiles in unprecedented detail.

\begin{figure}
\plotone{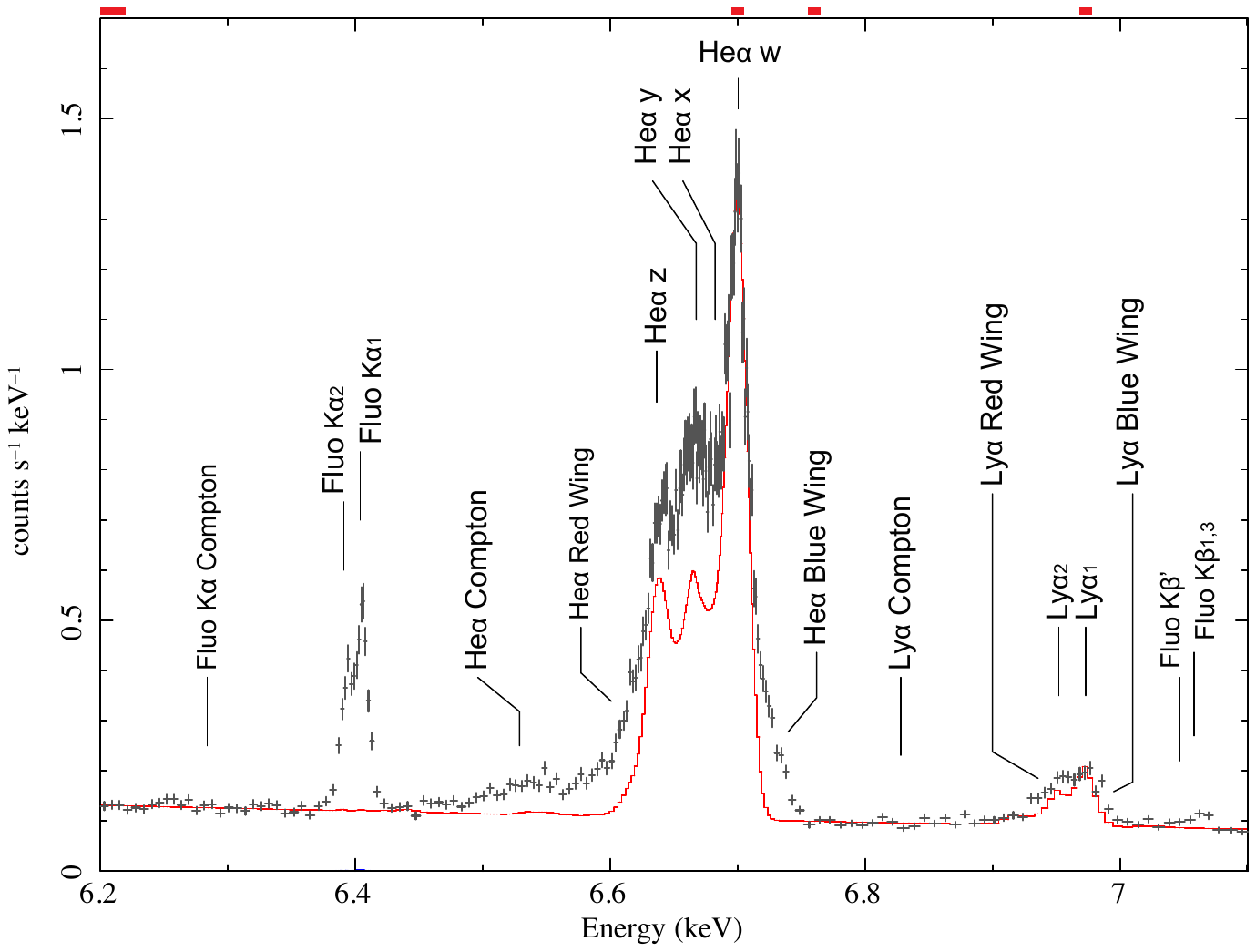}
\caption{\rsl Fe K$\alpha$ band spectrum of \etacar in XRI240609.
The solid red line shows the best-fit optically thin, collisional equilibrium one-temperature thermal plasma ({\tt bapec}) model
derived from a fit to the data in the top red bar ranges.
The blue line barely above the x-axis shows the estimated non-X-ray background spectrum.
\label{fig:spec620710_1Tfit}
}
\end{figure}

This paper focuses on the spectral features between 6.2 and 7.1 keV, which contain 
several Fe K$\alpha$ lines from various ionization states (Figure~\ref{fig:spec620710_1Tfit}).
We scrutinized the XRI240609 spectrum, which has an order of magnitude higher photon statistics than XRI231123,
and apply the result to XRI231123 for a comparison.

\subsection{Iron K$\alpha$ Line Profile}
\label{subsec:iron_Ka_lines}

The most conspicuous feature in this energy band is the Helium-like Fe K$\alpha$ (Fe He$\alpha$) line complex between 6.6$-$6.7~keV.
The narrow peak at $\sim$6.70~keV corresponds to the resonance line 
(He${\alpha}$-$w$).
It does not show a significant energy shift from the rest frame,
but it is significantly broader than the instrumental broadening.
The complex also has two peaks at $\sim$6.665~keV and $\sim$6.64~keV, 
which correspond to the inter-combination lines (He${\alpha}$-$y$) and the forbidden line (He${\alpha}$-$z$) \citep[e.g.,][]{Porquet2010}.
\rsl has enough spectral resolutions to separate these lines from each other,
but the intrinsic Doppler broadening blends them. The broadening also should submerge 
weaker emission lines between them.

The spectrum also shows two weaker peaks at $\sim$6.97~keV,
which correspond to the Hydrogen-like Fe K${\alpha}_{1}$ (Ly$\alpha_{1}$) 
and K${\alpha}_{2}$ (Ly$\alpha_{2}$) lines.
These lines are also heavily blended due to Doppler broadening.
The relatively sharp line at $\sim$6.40~keV corresponds to the fluorescent Fe K$\alpha$ line emission from quasi-neutral ionization states.
It also appears to have two peaks, whose energy difference is consistent with the K$\alpha_{1}$ and K$\alpha_{2}$ emission lines.
The spectrum also clearly shows a weak peak at $\sim$7.06~keV, corresponding to the fluorescent iron K$\beta$ line emission.

The spectrum also reveals structures that are not recognized in conventional stellar X-ray spectral profiles.
To highlight them, we compare the spectrum with a simple one-temperature plasma emission model,
which satisfies the spectrum in the following four specific energy bands:
Band $i$) 6.700$\pm$0.05~keV, including the broadened He$\alpha$ resonance line;
band $ii$) 6.973$\pm$0.05~keV, covering the Ly$\alpha_{1}$ line peak; and
bands $iii$) and $iv$) 6.2$-$6.22~keV and 6.755-6.765~keV dominated by electron-continuum emission without any strong emission lines.
These bands are depicted as the red bars in Figure~\ref{fig:spec620710_1Tfit}.
We fit these bands with a one-temperature velocity-broadened thermal plasma emission model ({\tt bapec}). 
Band $i$) mainly constrains the redshift ($z$) and the broadening velocity ($v_{\sigma}$).
The ratio of band $ii$) to band $i$) constrains the plasma temperature (\KT).
Bands $iii$) and $iv$) mainly constrain the continuum level --- the {\tt bapec} model normalization ($norm$).
The ratio of the line flux over the continuum constrains 
the elemental abundance ($Z$).
The best-fit parameters are 
\KT:~3.97~keV,
$Z$: 0.403~solar,
$z$: 0.00,
$v_{\rm \sigma}$: 332~\UNITVEL, and
$norm$: 0.126.
The best-fit model is plotted in red in Figure~\ref{fig:spec620710_1Tfit}.
We note that this model is designed to highlight the intrinsic spectral shape from the spectral response profile: 
the derived parameters, in particular the elemental abundance, should not necessarily reflect the actual plasma conditions.

The best-fit model reproduces the broadening of the resonance line profile fairly well.
The plasma temperature, which mainly represents the characteristic ionization temperature 
of highly ionized Fe,
is consistent with the slope above 7.1~keV, which traces the electron temperature, and
similar to the typical hottest plasma temperatures of {\etacar}'s WWC X-ray emission measured
in earlier observations \citep{Hamaguchi2007b,Hamaguchi2014b,Hamaguchi2016a}.
However, the model fails to reproduce the blue and red wings of both the He$\alpha$ and Ly$\alpha$ lines.
Lithium-like dielectronic recombination (DR) lines from cooler plasma may explain a fraction of the He$\alpha$ red wing, 
but no emission line explains the other wing excesses.
The model also explains only $\sim$60\% of the satellite line band flux between 6.63$-$6.69~keV.
Plasmas in collisional ionization equilibrium (CIE) at lower temperatures or 
non-equilibrium ionization (NEI) exhibit stronger emission in this band, but have a significantly 
weaker Ly$\alpha$ line.
These results suggest the presence of additional thermal components with substantial Doppler broadening and/or other mechanism.

The He$\alpha$ red excess extends down to $\sim$6.45~keV with 
a small peak at $\sim$6.54~keV.
To reproduce this excess by a Doppler-broadened He$\alpha$ Fe line would require a plasma velocity up to $v \approx$12000~\UNITVEL, 
which is much higher than those observed from \etacar or wind from 
any other massive stellar system.
On the other hand, cold materials that produce fluorescence Fe lines
can also Compton-scatter X-ray emission via electron interaction.
Compton scattering removes a fraction of the photon energy, 
depending on the scattering angle. 
He$\alpha$-$w$ photons at 6.70~keV down-scatter to 6.53~keV 
for 180\DEGREE\ back-scattering, consistent with the observed local peak.
The peak is not as clear as the source thermal spectrum,
because of multiple-angle scattering emission.
This excess emission should be the so-called Compton shoulder profile 
of the Fe He$\alpha$ line.
This would be the first clear detection of a Compton shoulder profile
associated with a thermal emission line, rather than the Fe K$\alpha$ fluorescence line.
In the meantime, the spectrum also shows a marginal enhancement above 
the continuum below the fluorescent K$\alpha$ line down to 
the back-scattering energy at $\sim$6.24~keV,
suggesting a detection of the Compton shoulder of
the Fe K$\alpha$ fluorescence line as well.

\subsection{Spectral Fitting Model}

Because the single-temperature model above provide a poor fit,
we add two additional Doppler-broadened components for the blue and red wings, fluorescent iron K$\alpha$ and K$\beta$ lines,
and Compton scattering components for each thermal component and the fluorescent K$\alpha$ line to the one-temperature model.
Hereafter, we call the initial {\tt bapec} component, which reproduces 
the relatively narrow He$\alpha$-$w$ line, the ``narrow" component and designate 
it as $N$.

A preliminary fit suggests that adding a single Gaussian broadened thermal component ({\tt bapec}) can reproduce most of the blue and red wing excesses.
However, 
lower-energy emission lines from Si, S, Ar, and Ca ions, 
which will be described in a subsequent paper, 
exhibit similar blue wings but significantly weaker or no red wings.
This suggests that the two wings have different origins 
and should be treated separately.
Reproducing a wing profile requires a smoothing only to one energy direction,
but the standard {\tt Xspec} library lacks such a convolution function.
We therefore develop an empirical asymmetric smoothing function, 
{\tt htsmooth},
which redistributes the flux in each spectral bin to a half-diagonal shape (Appendix~\ref{app:htsmooth}).
For each wing component, we convolved this function to an {\tt apec} 
model, whose temperature, abundance, and redshift parameters were tied 
to the corresponding parameters of the narrow component.

{\tt Xspec} includes several Compton scattering models, which, however, assume a power-law source spectrum.
We need a new model that considers down-scattering of thermal emission lines.
Compton scattering spectra can change with multiple conditions:
i) the emitting source's spatial and velocity distribution; 
ii) the reflector’s spatial and velocity distribution;
iii) the reflector's density, elemental abundances, and ionization states; and
iv) the orientation of the reflector to the observer.
A complete reproduction of the Compton spectral shape requires a detailed modeling of the WWC geometry,
which is beyond the scope of this paper.
We therefore developed an empirical Compton down-scattering model, {\tt emcomp} (Appendix~\ref{app:emcomp}), which assumes 
a simple redistribution function within the Compton down-scattering energy range and varies the function parameters to fit the Compton scattering profile. 
The current version does not account for photoelectric absorption so it is only applicable to 
restricted bands with negligible variation in absorption cross-section. 
We can safely assume that the three thermal components 
(narrow, blue wing, and red wing) are the sources of the Compton-scattered photons.
The main reflector is likely the primary wind on the opposite side of the emitting source from us during the observation,
and therefore sees the opposite plasma motion.
(see the discussion in section~\ref{subsec:geometry_wwc}).
We thus invert the redshift values of the thermal components.
This approximation over-simplifies the geometry of the reflector 
and observer, but 
still should reasonably describe the scattered emission, 
as the Compton down-scattering energy is considerably larger than 
the Doppler shift in the \etacar\ spectra.

\begin{figure}
\plottwo{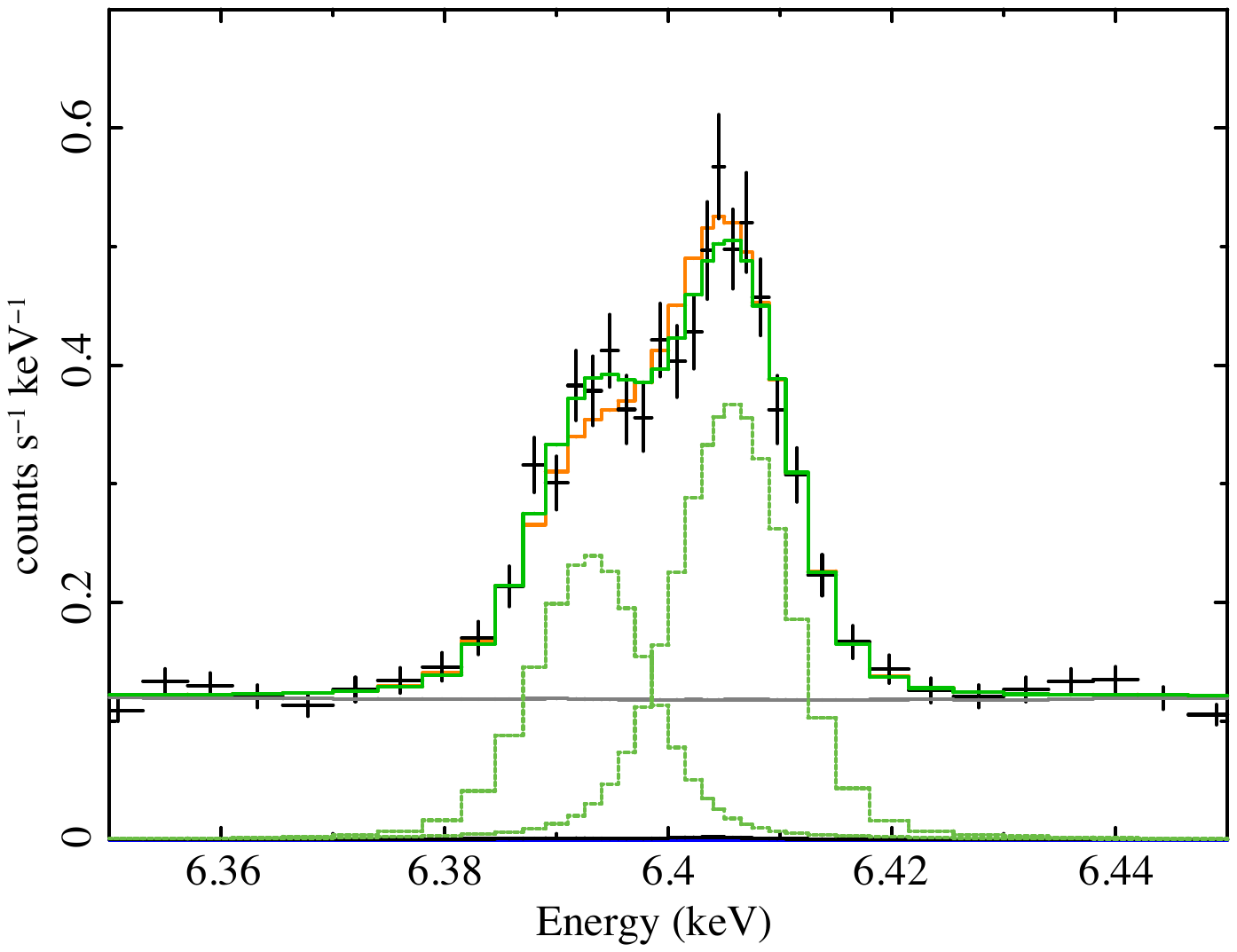}{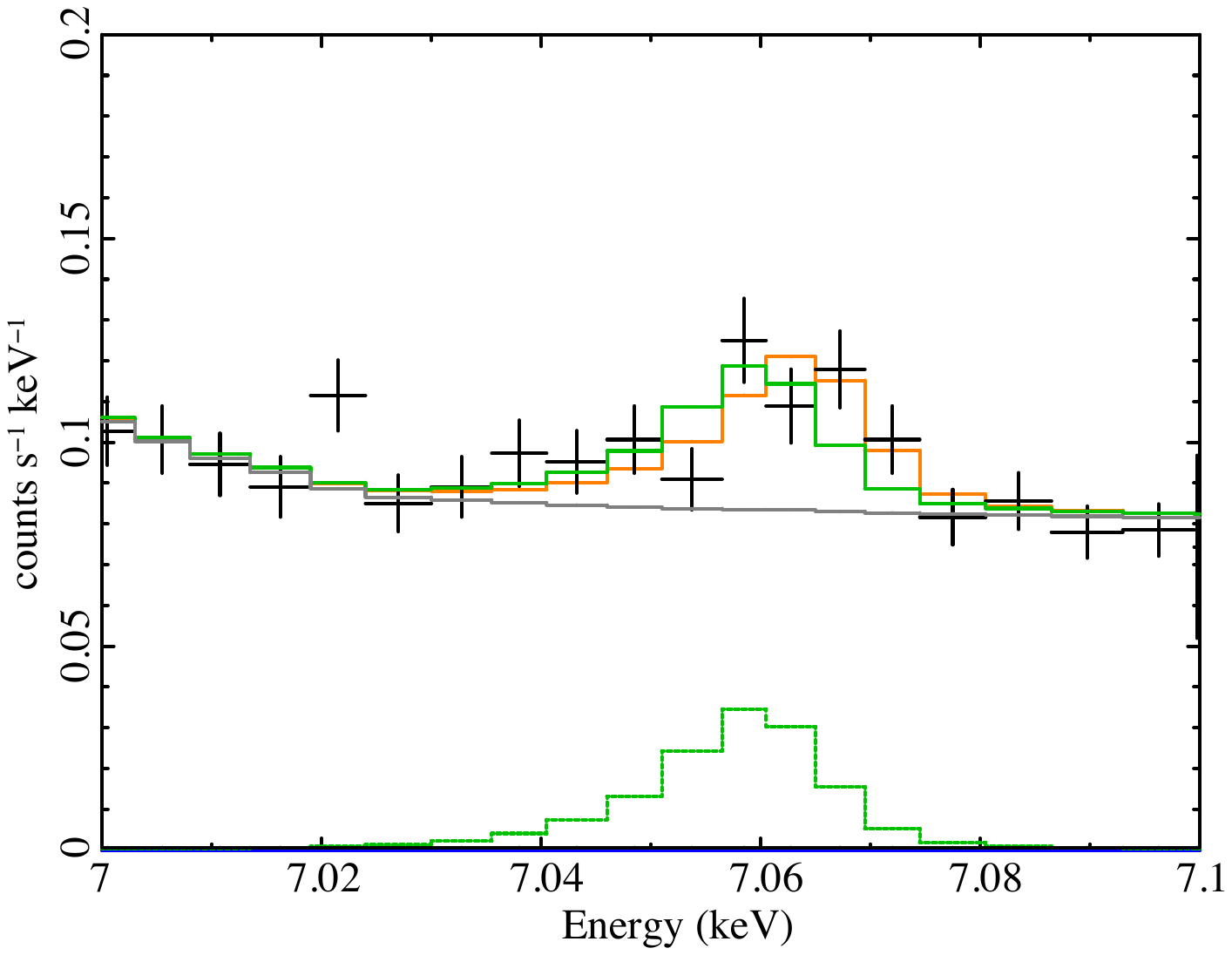}
\caption{Fluorescent Fe K$\alpha$ ({\it left}) and K$\beta$ ({\it right}) line spectra ({\it black}).
The line components are fitted by empirical models in \citet{Holzer1997a}, smoothed with the {\tt gsmooth} Gaussian convolution function.
For the K$\alpha$ lines, the original model exhibits a non-negligible residual 
between the K$\alpha_1$ and K$\alpha_2$ lines ({\it orange}).
We thus allow their line normalizations to be independently varied 
({\it green}).
For the K$\beta$ lines, 
we employ a model that ties the redshift and $v_{\sigma6}$ 
smoothing parameters to those of the K$\alpha$ model ({\it green}).
However, the best-fit model shows a marginal excess on the blue side,
and a model with a free redshift parameter reproduces this excess better
({\it orange}).
We employ the models in green for the entire model fit.
The green dotted lines shows the line components,
the grey lines represent a thermal model for the continuum, 
and the blue lines barely above the x-axis represent 
the non-X-ray background.
\label{fig:specFe64}
}
\end{figure}

For the fluorescent Fe K$\alpha$ and K$\beta$ lines, 
we refer to the model in \citet{Holzer1997a}, which reproduces the line shapes using multiple Lorentzian 
functions\footnote{The formulae are equivalent to the {\tt feklor} model series in {\tt Xspec}.}.
The observed emission lines are significantly broadened, so we
smooth the model using the Gaussian convolution function, {\tt gsmooth} in {\tt Xspec}, fixing the $\alpha$ parameter to 1 with the assumption that the lines are broadened by the same velocity.
A fit to the K$\alpha$ line spectrum with an {\tt apec} thermal emission for the continuum
moderately reproduces the line profile with a redshift 
at $-$3.00$\times$10$^{-4}$, equivalent to 
$-$89.9~\UNITVEL,\footnote{\label{nobarycorr}Observed velocity, 
not converted to the barycentric system.}
and smoothing sigma at 6 keV ($\sigma_{\rm 6}$) at 3.87~eV (193~\UNITVEL).
However, the best-fit model shows a small but systematic residual between 
the K$\alpha_1$ and K$\alpha_2$ lines from the \citet{Holzer1997a} model,
which assumes a 2-to-1 theoretical K$\alpha_1$$-$K$\alpha_2$ flux ratio
(Figure~\ref{fig:specFe64}, {\it left}, orange model).
A model that independently varies the K$\alpha_1$ and K$\alpha_2$ line normalizations eliminates this residual (Figure~\ref{fig:specFe64}, {\it left}, green model),
so we use this model for the entire spectrum fit.
For the fluorescent Fe K$\beta$ line, we tied the redshift and {\tt gsmooth} sigma parameters to those of the K$\alpha$ line, but fit the model normalization independently (Figure~\ref{fig:specFe64}, {\it right}, green model).
We note that the best-fit model leaves a marginal residual on the blue side, which can be reproduced more accurately with 
an independent redshift value at $-$9.70$\times$10$^{-4}$ 
($-$291~\UNITVEL,\footnote{\ref{nobarycorr}}, Figure~\ref{fig:specFe64}, {\it right}, orange model).

The Compton shoulder of the Fe fluorescent K$\alpha$ line is weak, so
we tied the redistribution function parameters to those of the Compton scattering of 
the thermal emission model, and found a good fit.
We note that the redistribution function can differ,
as the emitting source geometry is distinct:
the primary wind for the Fe fluorescence Compton scattering,
and the WWC thermal plasma on the WWC shock cone 
for the thermal Compton scattering.

\subsection{Fitting the XRI240609 Spectrum}

We combined all these components into a model and added the best-fit NXB background to the spectral model (see Appendix~\ref{app:nxb}).
We fit the spectrum between 6.2$-$7.1~keV with this model using C-statistics \citep{Cash1979a}.
We estimated the 90\% confidence ranges of the fitting parameters using a Markov Chain Monte Carlo (MCMC), Goodman-Weare algorithm 
built in {\tt Xspec} {\tt chain}.
With 10 walkers, we produced five sets of 10,000 data lengths after burning 200 data lengths.
The redshift parameter of the thermal components in XRI240609 was not constrained appropriately, possibly due to an anomalous model behavior near zero.
Since the He$\alpha$-$w$ line profile mainly constrains the redshift parameter,
we evaluate a 90\% confidence range for the one-temperature {\tt bapec}
model from the 4 band spectrum, after fixing the $v_{\sigma}$ parameter
to the best-fit value.

\begin{figure}
\plotone{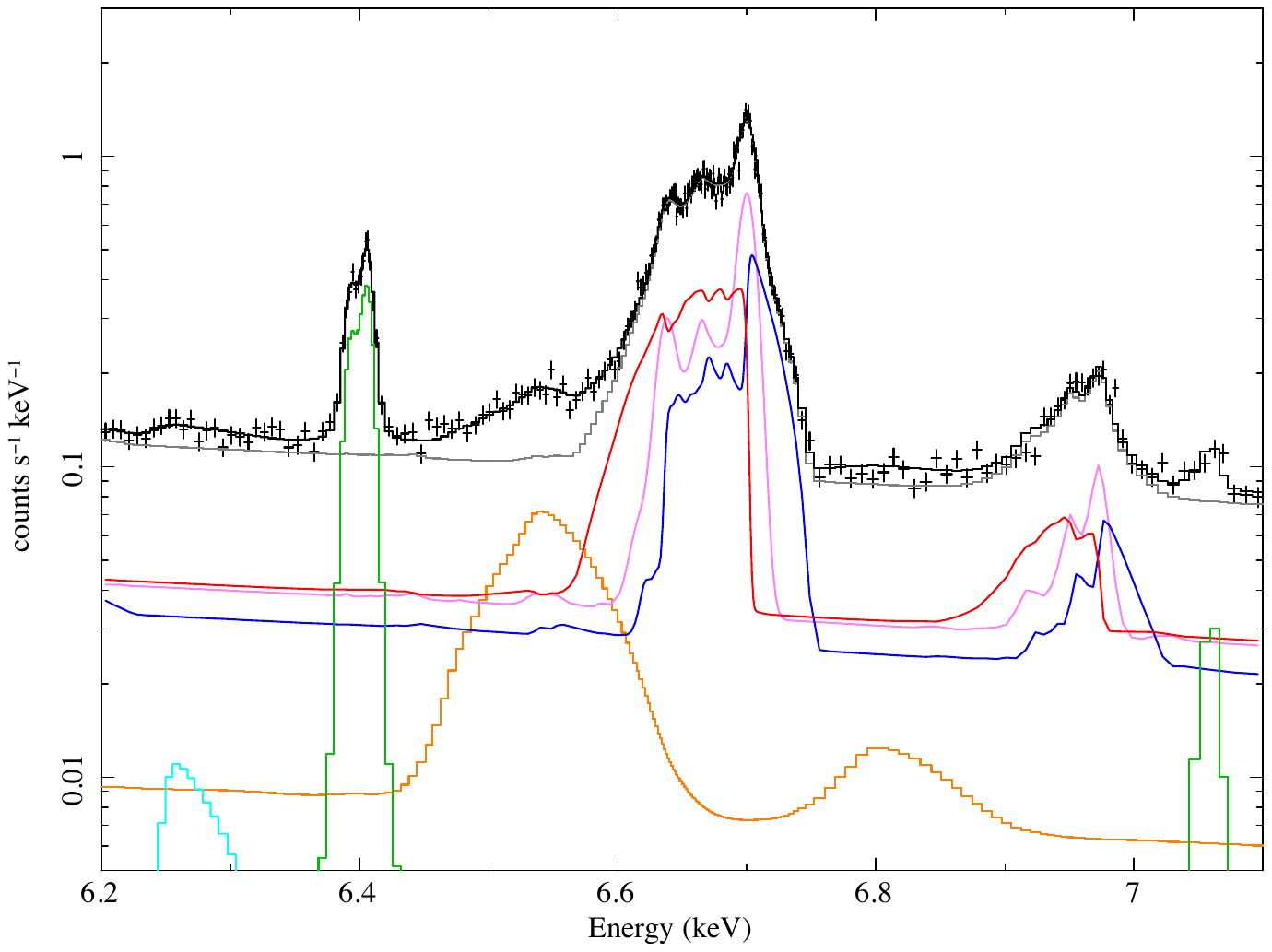}\\
\vspace*{-2cm}
\plotone{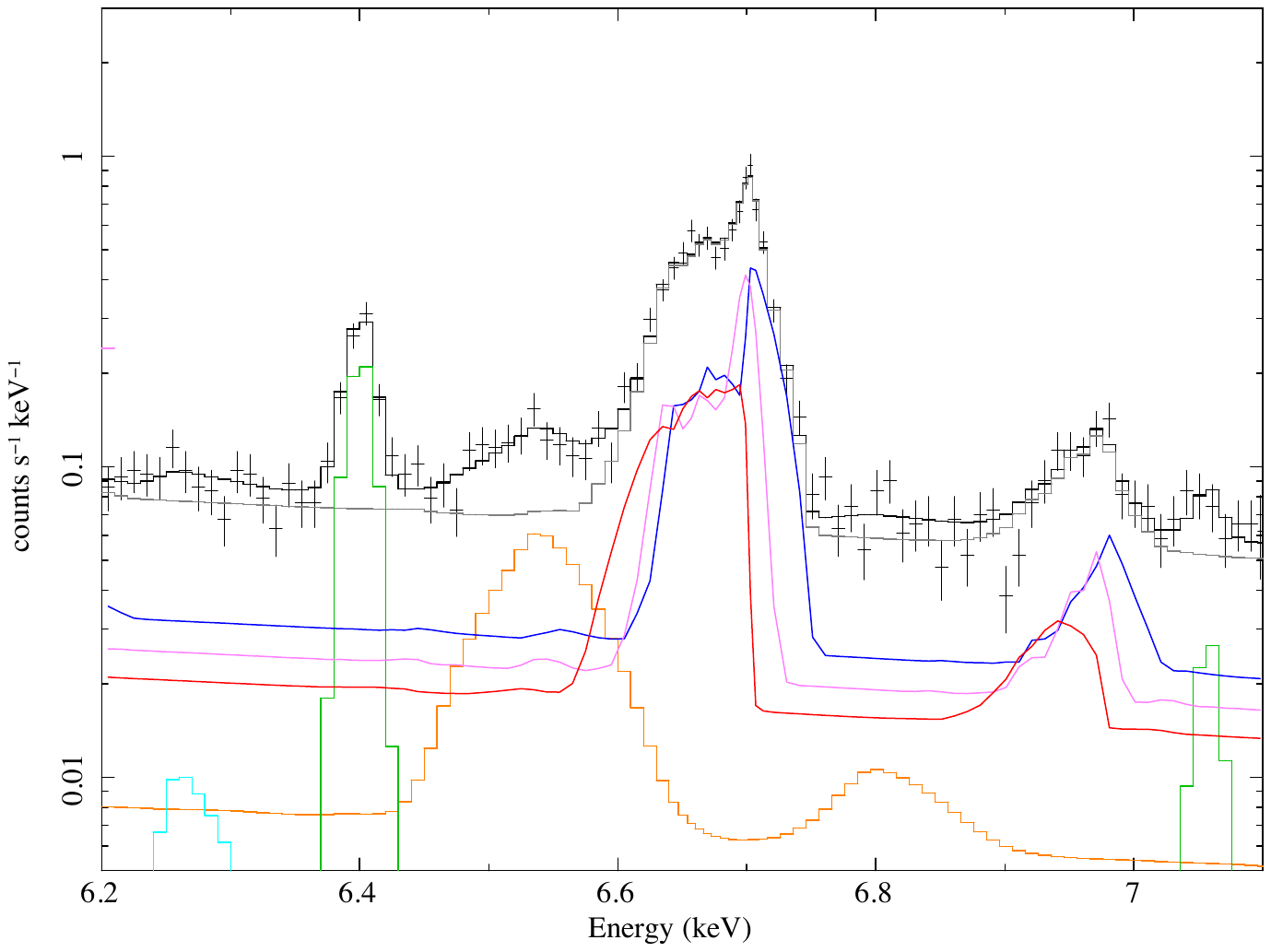}
\vspace*{-1cm}
\caption{
Best-fit model of the Fe K$\alpha$ band spectrum between 6.2$-$7.1~keV ({\it top}: XRI240609, {\it bottom}: XRI231123).
The black and grey lines represent the total and the thermal plasma emissions. 
The blue, purple, and red lines show the blue wing, narrow, and red wing components, and the orange line 
represents the Compton scattering of these components.
The green line shows the Fe fluorescent K$\alpha$ and K$\beta$ lines, and the cyan line shows the Compton scattering of the K$\alpha$ line.
The non-X-ray background background, included in the model, is below this plotting scale.
\label{fig:spec620710_bestfit}
}
\end{figure}

\input{tab2}

The best-fit model reproduces all the spectral profiles very well (Table~\ref{tbl:bestfitspec}, Figure~\ref{fig:spec620710_bestfit}).
Three plasma components with a similar brightness, having \KT~=4.06~keV and $Z =$0.71~solar, 
reproduce both the He$\alpha$ and Ly$\alpha$ lines, including the satellite lines.
The narrow component has a moderate Doppler broadening of 
$v_{\sigma}^{\rm N}$~$\sim$293~\UNITVEL\ but no significant radial velocity.
The blue wing component has a maximum blueshift of $v_{\rm max}^{\rm BW} \sim-$2220~\UNITVEL, 
while the red wing component has an even higher maximum absolute velocity of $v_{\rm max}^{\rm RW} \sim$3107~\UNITVEL.
These results are consistent with the picture that the thermal emission originates from 
collisional equilibrium plasmas streaming from the WWC apex, which moves at a small ($<$ 100~\UNITVEL) radial velocity.

A Compton scattering spectrum reflecting 8\% of the thermal emission reproduces the excess below the He$\alpha$ red wing,
including the peak at $\sim$6.54~keV.
The {\tt emcomp} redistribution function (Appendix~\ref{app:emcomp}) is nearly triangular, 
with the smallest scattering fraction (SSF) at nearly zero.
The smallest scattering cut (SSC) value, 0.38, is equivalent to 6.63 keV 
for the He$\alpha$-$w$ line,
where the He$\alpha$ red wing is relatively strong.
Since both the smoothing and Compton reflection functions are empirical,
the derived SSC value strongly depends on the assumed red wing model,
and should be interpreted with caution.

The fluorescent iron lines are significantly blueshifted at
$\sim-$126~\UNITVEL\ 
and Doppler broadened by 
$\sim$181~\UNITVEL.
The K$\beta$ line accounts for 8.9\% of the flux of the K$\alpha$ line,
which is marginally lower than expected from a laboratory measurement \citep[11.3\% of the K$\alpha$ line,][]{Thompson2009book}.

\subsection{Fitting The Commissioning Phase Observation, XRI231123}

XRI231123 shows a very similar \rsl spectrum to XRI240609 
between 6.2$-$7.1~keV, except that the count rate is $\sim$68\% of 
that of XRI240609.
We therefore used the same spectral model to fit the XRI231123 spectrum.
The spectrum lacks sufficient photon statistics to reproduce the Fe fluorescence and Compton shoulder profiles accurately.
Meanwhile, the profiles should not vary significantly between the observations with similar shock cone orientations.
We, therefore, fixed the Fe fluorescent K$\alpha$ line ratio and the SSC and SSF values
to the best-fit values of XRI240609.
Table~\ref{tbl:bestfitspec} and Figure~\ref{fig:spec620710_bestfit} 
display the best-fit parameters.

\section{Discussion} \label{sec:disc}

\subsection{Origin of the Three Thermal Components}

The \XRISM\ \rsl observations detect
three radial velocity components in the Fe K$\alpha$ thermal plasma emission lines:
a narrow component with a Doppler broadening of $v_{\sigma}^{\rm N} \sim$300~\UNITVEL,
and two wing components with $|v_{\rm max}| \sim$2100$-$3100~\UNITVEL.
\CHANDRA/{\it HETG} observations,
which have previously provided the highest-resolution spectra in this energy band, could measure the average radial and broadening velocities of the He$\alpha$ line 
but lacked the spectral resolution to separate these velocity components. 
The \XRISM\ \rsl spectrum of \etacar clearly demonstrates the superb energy resolution and efficiency of 
the \XRISM\ \rsl spectrometer in the Fe K$\alpha$ band.

\begin{figure}
\plotone{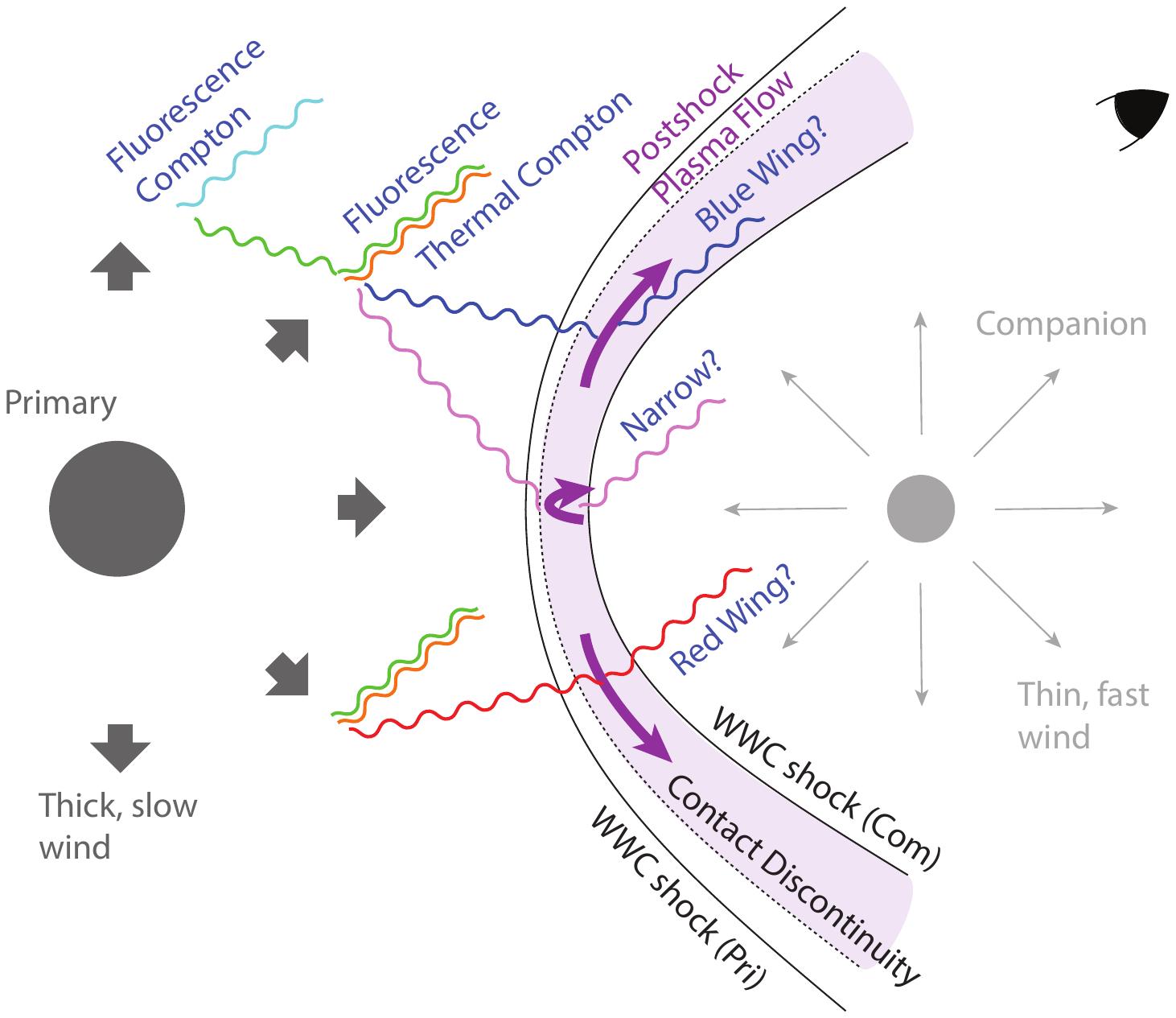}
\caption{
Schematic view of the shock geometry and X-ray emitting locations in the current model.
The WWC produces two shocks, one on each side of the wind contact discontinuity.
The observed X-ray emission would arise from the post-shock plasma of the companion wind, which is depicted in filled purple.
The heated plasma flows down along the WWC contact discontinuity, 
forming the blue and red wings depending on 
the flow direction relative to the viewing angle.
Some plasma may stand near the stagnation point 
and emit the narrow component. 
We observe these X-ray emissions through the cavity produced 
by the companion wind.
Iron atoms in the primary wind absorb the WWC plasma emission above 7.1~keV 
and re-emit fluorescent lines, 
while electrons in the wind Compton-scatter the WWC emission.
These electrons can also Compton-scatter fluorescing X-rays.
\label{fig:EmitPath}
}
\end{figure}

The primary wind with a velocity of 420~\UNITVEL\ can thermalize gas only up to 
$\sim$0.3~keV \citep[e.g., equation (1) in][]{Luo1990a}.
The highest temperature can be even lower as the radiative cooling
should work efficiently for the post-shock primary wind 
\citep[see][]{Madura2013}.
The observed \KT~$\sim$4~keV plasma emission should originate from 
the post-shock secondary wind with $v \sim$3000~\UNITVEL\ and 
\Mdot\ $\sim$10$^{-5}$~\UNITSOLARMASSYEAR\ \citep[e.g.,][see Figure~\ref{fig:EmitPath}]{Corcoran1998a,Pittard2002}.

The wing components have maximum broadening velocities similar to 
the terminal wind velocity of the companion and far above 
that of the primary wind.
This result again suggests that the \KT~$\sim$4~keV plasma 
originates from the post-shock companion wind.
However, the pre-shock wind decelerates with the shock:
the post-shock velocity of the head-on collision at the WWC apex
should be nearly zero.
The post-shock flow downstream adds momentum from oblique collisions 
and accelerates \citep[see][]{Miyamoto2024a}.
The high-velocity plasma may have already accelerated 
at relatively far distances from the WWC apex.

If the downstream plasma flows axis-symmetrically through the contact discontinuity,
the velocity profile should have a double-horn structure with peaks near the
high and low velocity ends \citep{Henley2003a}.
However, the observed line profile has a triangular shape with a peak near the center.
This may suggest that the observed line profile is not produced solely by a laminar 
flow of the WWC downstream, but rather by another broadening mechanism, such as turbulence. 
We note that observed X-ray line profiles from WR 140, a well-known colliding-wind binary system, also exhibit triangular profiles \citep{Miyamoto2024a}. 

The Fe fluorescence and Compton scattering profiles suggest that 
the companion star is on our side during these \XRISM\ observations
(see Section~\ref{subsec:geometry_wwc}).
In this orientation,
the downstream flow should be mainly blueshifted and therefore 
the blue wing component should be more luminous with a higher absolute maximum velocity 
than the red wing component. 
However, the observed red wing has a higher absolute maximum velocity, 
and a higher flux, at least in XRI240609.
This may suggest that the opening angle of the WWC shock cone is near 180\DEGREE.
If so, the companion star has a significantly higher wind velocity and/or mass loss rate 
than previous estimates, or the primary star has a lower mass loss rate.

The narrow component has a distinct velocity profile from the wing components.
It has nearly zero radial velocity at $\sim$ $-$12.8$\pm$27~\UNITVEL\ 
and relatively small broadening velocity, $v_{\sigma}^{\rm N} \sim$290~\UNITVEL,
suggesting that it originates from a slower part of the WWC.
\XRISM\ \rsl spectra of another WWC binary, WR~140, 
whose two stars have thin, fast winds ($V_\infty\sim$3000~\UNITVEL),
only have broad Fe He$\alpha$ line components of $|v_{\rm max}| \sim$3000~\UNITVEL\ 
(Hamaguchi et al., private communication).
The difference may suggest that the thick, slow primary wind of \etacar
plays a significant role in producing the narrow component.

Although {\etacar}'s binary orbital solution is not yet known,
the WWC apex should have a low velocity along with the orbital motion
during the observations.
The optical H$\beta$ line, which should trace the primary wind,
has radial velocities of 0$-$10~\UNITVEL\ \citep{Strawn2023a}.
This implies that the WWC apex radial velocity is within $\pm$10~\UNITVEL, 
assuming the presumed stellar mass ratio of $\sim$2$-$3 and wind momentum ratio of 0.12$-$0.42.
The redshift of the narrow component is consistent with the WWC apex's radial motion.
This narrow component may arise from the shock-stagnation point of 
the companion wind, where the relative flow velocity is nearly zero.
However, the stagnation point is theoretically a point, and it is not clear 
if such a low velocity region can, in reality, be large enough to produce 
emission comparable to each high-velocity wing component.
Alternatively, a part of the post-shock companion wind near the apex may penetrate 
the primary wind and interact with the relatively cold primary wind.
This hypothesis may explain why the broadening velocity is close to the primary wind velocity and the WR~140 spectra do not show a similar narrow component.

\subsection{Geometry of the WWC Shock}
\label{subsec:geometry_wwc}
\label{subsec:disc_fluorescent_iron_K}

The observed fluorescent Fe K$\alpha$ lines are significantly blueshifted compared to those from neutral Fe at the rest frame.
Although the line energies change with ionization, 
most iron atoms in the cold primary wind do not exceed the charge state Fe$^{3+}$ \citep[e.g.,][]{Madura2013},
and the line centroid energies remain unchanged within this ionization range \citep{Yamaguchi2014a}.
The observed blueshift reflects the radial velocity of the fluorescing matter rather than ionization.

About 90\% (1.62$\sigma$) of the fluorescent photons in XRI240609 are
within the primary wind's radial velocity range between 
$-$420~\UNITVEL\ and +168~\UNITVEL.
The lower end, $-$420~\UNITVEL, is the blueshifted terminal velocity 
of a stream of the primary wind flowing toward us.
The higher range between +168 and +420~\UNITVEL\ is absent, plausibly due to 
a thick optical depth to the far side of the primary wind.
This result strongly suggests that the fluorescent Fe lines 
mainly originate in the primary wind viewed from a side of the companion.
This geometry may also explain the apparent K${\alpha}_2$/K${\alpha}_1$ flux 
ratio anomaly: a fluorescing line from an optically thick, spherical wind 
at an oblique angle to the irradiating source produces
an asymmetric line profile with a red tail.

Compton scattering occurs in any medium with electrons, free or bound, 
and should happen on the paths that produce Fe K fluorescence as well.
The best-fit {\tt emcomp} distribution function has a triangular shape with
more backscattering photons, again indicating that the majority of 
the scattering medium lies behind the emitting source, i.e., the WWC plasma, 
as seen from us.
The best-fit result shows little scattering below SSC = 0.38, 
corresponding to a scattering angle of $\approx$76\DEGREE, 
suggesting little scattering matter along the line of sight and 
the wind half-opening angle at $\approx$76\DEGREE\ or larger.
It is larger than the half-opening angle of the WWC 
shock cone of 55$-$75~\DEGREE\ expected from the wind velocities 
and mass loss rates \citep[e.g.,][]{Madura2013}, but largely 
consistent considering a hidden uncertainty of the SSC measurement.

The best-fit Compton Flux Ratio (CFR) suggests that 
$\sim$8\% of the source emission is reflected with Compton scattering.
Since 80\% of the photons on the scattering paths,
assuming the elemental abundance discussed in Section~\ref{subsec:disc_Compton_shoulder},
suffer photoelectric absorption,
the primary wind as scattering matter should subtends $\sim$1.6$\pi$ steradians 
from the emitting source, which is equivalent to the companion wind's half-opening angle at $\sim$110\DEGREE, if it is optically thick.
Since some scattering paths away from the primary star would be optically thin,
this provides an upper limit of the wind opening angle, 
consistent with the above constraint.

All these results indicate that 
the primary wind is on the opposite side of the WWC apex from us, 
i.e., the companion star is on our side, during the observations.
This orientation is consistent with the several orbital models constrained 
from optical or X-ray emission line profiles \citep[e.g.,][]{Strawn2023a,Russell2016a,Madura2013,Henley2008}
but not the models based on the assumption that 
the absorption increase near periastron originates from 
mass accretion onto the companion star \citep[e.g.,][]{Kashi2021a,Kashi2009a}.

\subsection{Elemental Abundance of the Primary Wind}
\label{subsec:disc_Compton_shoulder}

Intensities of the Compton-scattered emission and the Fe fluorescent line are proportional to the number of electrons and Fe atoms, respectively,
so their flux ratio constrains the elemental abundance of the primary wind.
To minimize the effects of absorption and the uncertainty in the {\tt emcomp}
distribution function, we evaluate the Compton-scattered emission at 6.4~keV, the Fe fluorescent line energy and use the line equivalent width (\EW), defined as the 
intensity ratio of the Fe K$\alpha$ fluorescence line (the K${\alpha}_1$ + K${\alpha}_2$ flux
in Table~\ref{tbl:bestfitspec}) to the Compton scattering continuum 
(the orange line in each panel of Figure~\ref{fig:spec620710_bestfit}) at 6.4~keV.

This measurement provides a robust estimate of the number of electrons per Fe atom,
which is $\sim(N$(H) + 2$N$(He))/$N$(Fe) as H and He are the dominant electron carriers.
H and He are key elements for understanding stellar evolutionary stages, 
but measuring their abundances using UV or optical emission lines is challenging 
due to inherent uncertainties in their ionization and chemical states.
This X-ray measurement is largely unaffected by the ionization or
chemical state of the primary wind, though it cannot count each element separately.

The {\EW}s of the best-fit models are $\sim$947 (73)~eV for XRI240609 and 834 (155)~eV for XRI231123,
where the values in parentheses indicate the 90\% confidence ranges.
We evaluate the expected {\EW}s for a 1 solar ({\tt aspl}) elemental abundance absorber irradiated by \KT\ =4 keV plasma emission.
First, we calculate these values numerically for an optically-thin absorber, using the Thomson scattering cross-section,
which averages out the angular dependence of Compton scattering, for simplicity.
At each photon energy $E$, the Compton scattered flux is {\tt SRC}($E$)${\times}N_{\rm e}{\times}{\sigma}_{\rm TH}$, 
and the flux absorbed by Fe atoms in the wind is 
{\tt SRC}($E$)${\times}N_{\rm Fe}{\times}{\sigma}_{\rm Fe}^{\rm abs}({\it E})$, with {\tt SRC}($E$) being the incident flux at energy $E$, $N_{\rm e}$ 
the column density of electrons, 
$N_{\rm Fe}$, the Fe column density, 
and ${\sigma}_{\rm TH}$ and ${\sigma}_{\rm Fe}^{\rm abs}(E)$
the Thompson scattering and Fe absorption cross sections, respectively.
For the \EW of the Fe K$\alpha$ fluorescence line, 
we integrate the flux absorbed by Fe atoms between 7.11~keV and 28~keV\footnote{
Note that more than 80\% of the fluorescent emission originate from incident photons between 7.11$-$10~keV}, 
and multiply this by the Fe fluorescence yield (35.1\%) and the flux ratio of the K$\alpha$ lines 
\citep[89.8\%,][]{Thompson2009book}.
Dividing the Fe K$\alpha$ integrated line flux by the Thomson scattered continuum at 6.4~keV, 
we obtain an \EW of 582~eV.
We also consider the more realistic case of an optically-thick absorber 
using the Monte Carlo radiative transfer code, SKIRT
\citep{VanderMeulen2023a,VanderMeulen2024a}, assuming a point-like, 4~keV thermal source seen through a flat, optically-thick slab \citep[similar to the geometry in section 4.1.2 of ][]{VanderMeulen2023a}.
The simulations indicate an \EW of 584 eV when electrons are bound to atoms,
and at 559~eV when electrons are unbound to atoms.
Both the numerical and simulation results suggest that the \EW does not depend significantly on the optical thickness or the ionization of the absorber.
The observed {\EW} is about 60\% higher than any of the calculated values assuming 
the {\tt aspl} solar abundance medium
\citep{Asplund2009}.

Contamination of the primary wind by CNO nuclear fusion products 
would primarily cause the electron depletion. 
The CNO cycle produces one He atom from four H atoms, diminishing 
the number of electrons by 2.
CNO processing also increases N and decreases C and O 
\citep{Davidson1982,Leutenegger2003}.
However, since CNO act as catalysts,
the total number of CNO nuclei remains constant, and therefore the total number of electrons in the nitrogen-enhanced gas
changes by less than 4\%.
We simulate how the H and He abundances change from a solar abundance gas due to nuclear fusion,
and calculate an {\EW} at each set of abundances,
assuming no C or O atoms, and 12.2 solar for N,
and abundances of other elements at 1 solar.
Then, the abundances of H at 0.16 solar and He at 3.47 solar, 
$N$(H)/$N$(He) =0.54, or the H mass ratio at $\sim$0.12,
provide an \EW at $\sim$948~eV, nearest to the observed value.

Such a low H abundance on the stellar surface is predicted 
in the late {\it LBV} or nitrogen-rich Wolf-Rayet (WN) phase
according to a stellar evolution model for a non-rotating 60~\UNITSOLARMASS\ star \citep{Groh2014a}.
Eta Carinae is expected to have a significantly larger 
initial mass of $\sim$150~\UNITSOLARMASS\ \citep{Hillier2001},
and therefore the result may not be directly applicable.
Still, the result may suggest that He-burning in the stellar core 
might already have started. 
In the meantime, the derived $N$(H)/$N$(He) ratio is significantly smaller than the assumption in \citet{Hillier2001} for constraining 
the mass of \etacar.
If this value is correct, the primary star's mass loss rate would be 
higher by a factor of $\sim$4.7  
and that the minimum total mass of the two stars would be smaller to $\sim$87~\UNITSOLARMASS\ \citep[see Sections~9.1 and 12 of][]{Hillier2001}.

An enhanced Fe abundance also results in a high \EW.
While most abundance tables have similar Fe abundances\footnote{see https://heasarc.gsfc.nasa.gov/xanadu/xspec/manual/node116.html} to the \texttt{aspl} table we adopt, the {\tt angr} abundance table  \citep{Anders1989} has a $\sim$48\% higher Fe abundance than 
the \texttt{aspl} table.
This Fe abundance yields a calculated \EW of 806~eV with 1 solar 
H and He abundances, which is 
still lower than the observed value. 
Note that the Fe and Ni emission lines from hot plasmas in the \rsl spectra are in better agreement with a model using the \texttt{aspl} table,
while the \texttt{angr} abundance table significantly overpredicts 
the Fe He$\alpha$ line strength relative to the Ni He$\alpha$ line.

\section{Conclusions} \label{sec:conclusion}

\XRISM\ observed the supermassive stellar binary system, \etacar,
which drives the most vigorous wind-wind collision X-ray activity within 3~kpc,
in 2023 November ($\phi_{\rm X}$ =0.68) and 2024 June ($\phi_{\rm X}$ =0.78).
The \rsl X-ray micro-calorimeter observations provided unprecedented 
energy resolution spectra of the binary system between 1.7$-$12~keV,
which clearly separate emission lines from major elements, Ni, Fe, Ca, Ar, S, and Si
as well as detect weak lines from rare elements such as Mn and Cr.
In particular, the spectra exhibit the detailed structure 
in the Fe~K$\alpha$ band between 6.2$-$7.1~keV for the first time,
revealing multiple velocity components in the
Fe He$\alpha$ and Ly$\alpha$ thermal emission lines,
Doppler motions of the fluorescent Fe K$\alpha$ and K$\beta$ lines,
and the presence of their Compton-scattered emissions
--- arguably the first clear detection of the Compton shoulder 
of thermal X-ray emission lines from celestial sources.

The Fe He$\alpha$ and Ly$\alpha$ thermal emission lines are significantly broadened, blending individual emission lines, 
including He$\alpha$ satellite lines. 
The broadening line profile requires three velocity components, narrow, blue wing and red wing, of \KT~$\sim$4~keV collisional equilibrium plasmas, 
but it does not require exotic states such as non-equilibrium plasmas. 
These plasmas have similar  characteristics to those observed in earlier 
\etacar\ at this orbital phase. 
The two wing components can be reproduced with asymmetric broadening shapes with maximum radial velocities at $\sim$2100$-$3200~\UNITVEL. 
These velocities are close to the terminal velocity of the companion wind, suggesting that these components originate from the post-shock companion wind flowing through the contact discontinuity. 
However, the observed triangular velocity distribution does not match the double-horn profile expected from a laminar flow through the shock cone:
it may require an additional broadening mechanism, such as turbulence within the flow.
In contrast, the narrow component has a negligible radial velocity with small broadening at $v_{\sigma}^{\rm N}\sim$293~\UNITVEL. 
These velocity profiles suggest that the plasma is trapped at the stagnation point at the WWC apex and/or penetrated into the primary wind.

The XRI240609 spectrum clearly separates the Fe fluorescent K${\alpha}_{1}$ 
and K${\alpha}_{2}$ lines at $\sim$6.4~keV and 
detects the Fe fluorescent K${\beta}$ lines.
These lines exhibit a significant blueshift by $\sim-$126~\UNITVEL\
and a broadening of $\sim$181~\UNITVEL\ 
with a marginal asymmetrical broadening structure.
The XRI231123 spectrum also shows a similar result with limited statistics.
Both spectra also exhibit He$\alpha$ Compton shoulder emission
between 6.45$-$6.60~keV clearly, as well as Compton shoulder profiles
associated with the Ly$\alpha$ and fluorescent Fe K$\alpha$ lines marginally.
The emission profile suggests strong backscattering emission
and little forward scattering emission.
Both the Fe fluorescent line dynamics and the Compton scattering
profile indicate that the absorbing and scattering medium is 
the primary wind on the opposite side of the WWC plasma from us,
i.e., the companion star is on our side, during the observations.

The flux ratio of the Fe fluorescent line over the Compton scattering
is significantly larger than that expected from a solar abundance 
absorber or scatter,
indicating that the primary wind is heavily depleted in electrons.
This phenomenon can be explained by the contamination of the primary 
wind with nuclear fusion products from the CNO cycle, which converts 
4 H atoms into 1 He atom.
Assuming that the Fe abundance does not change with stellar evolution,
the observed ratio requires a very low H abundance of 
$\sim$0.16 solar, which can be realized in a late {\it LBV} phase in 
a stellar evolution model.

These unprecedented details of the \XRISM/\rsl\ Fe K-band spectra  
provide valuable insights into the conditions of both hot and cold gases,
but also introduce further mysteries.
Addressing those questions requires more in-depth analyses of the data, including emission lines 
beyond the 6.2$-$7.1~keV range.
This includes examining Si, S, Ar and Ca lines between 2$-$5~keV, 
as well as Ni and Fe He$\beta, \gamma$ lines beyond 7.1~keV,
and utilizing more advanced modelings.
Additionally, some perplexing issues, such as the ``narrow" component, will necessitate further \XRISM observations of \etacar\ 
during different binary orbital phases and 
other massive WWC stellar binary systems.

\input{acknowledgment}

\facilities{XRISM(Resolve, Xtend)}

\software{
      HEASoft \citep{Heasarc2014a},
      xspec \citep{Arnaud1996},
      scipy \citep{SciPy2020a},
      astropy \citep{Astropy2013a},
      AtomDB ver. 3.1.2 \citep{Foster2012a}
      }


\appendix
\section{Gain recalibration of the XRI231123 data}
\label{app:gaincal}
This observation was performed during the commissioning phase of the \XRISM\ observatory when many of the methods used for calibrating the instruments were developed. The \rsl\ instrument, in particular, requires careful calibration of the time-dependent energy scale during each observation. The time-dependent energy scale is dependent on extrinsic factors including the heat sink temperature for the 50~m\DEGREEKELV\ focal plane, the amount of bolometric loading, and the temperature of the instrument electronics which affects the gain and bias current for the detector system. The original plan was to use the on-board modulated X-ray sources (MXS) to periodically perform this calibration during each observation \citep{Sawada2024a}. Unfortunately, the instrument gate valve did not open as planned, and the MXS could only illuminate about half of the array. Thus the current strategy is to use $^{55}$Fe radioactive X-ray sources mounted on the instrument's filter wheel (FW) to periodically provide fiducial measurements of the gain, and thus the energy scale, of the instrument. The periodic measurements of the 5.9~keV Mn K${\alpha}$ from the FW is used linear interpolate between fiducial measurements and to synthesize the non-linear gain scale for each X-ray event \citep{Porter2016a,Porter2024a}. Unfortunately, this observation was performed before this process was established.

\begin{figure}
    \plotone{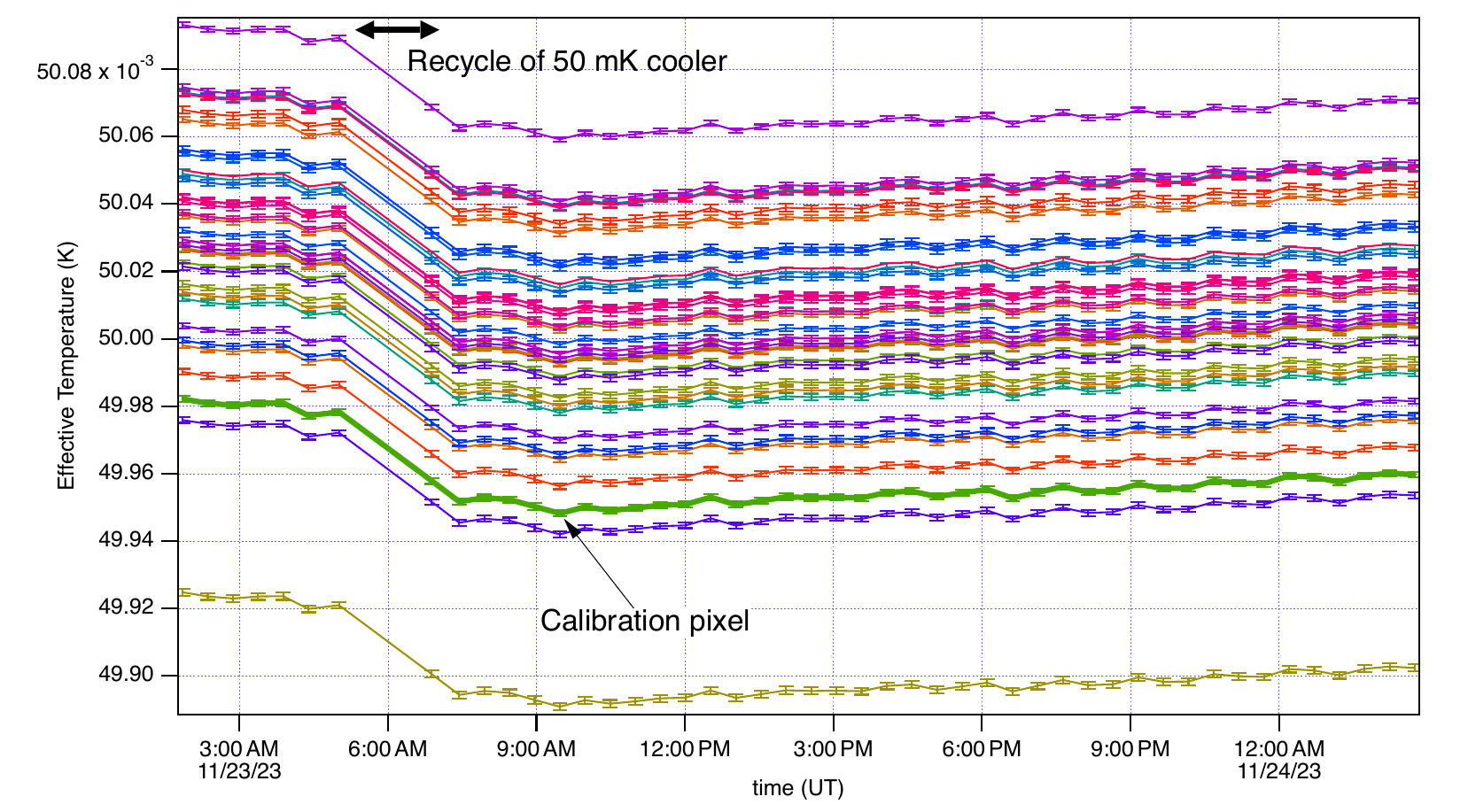}
    \caption{The time dependent energy scale (detector gain) vs time at 5.9~keV synthesized from the last known fiducial measurement and a continuous measurement using the calibration pixel (heavy green). The detector gain is parameterized as "effective" pixel temperature \citep{Porter2016a} and each pixel is a different color.
    \label{fig:gain}
    }
    \end{figure}

For this observation, the closest fiducial measurement was about two days before the observation, and there were no fiducial measurements during the observation. Instead, we used the last fiducial measurement for each pixel and extrapolated the gain of each pixel forward using a calibration pixel located outside of the field of view of the instrument. The calibration pixel is part of the focal plane array but located off to the side of the instrument field of view. It is continuously illuminated by a heavily collimated dedicated $^{55}$Fe X-ray source. The reason that the calibration pixel is not the sole gain calibrator for the entire array is due to differential bolometric loading between pixels in the focal plane and time-dependent changes in the detector bias that can cause the gain of the pixels to diverge over long periods of time. However, during short intervals, such as the two days between the last fiducial measurement and this observation, the divergence is small, likely better than 1~eV at 6~keV. Figure~\ref{fig:gain} shows the time dependent energy scale for this observation for each pixel as synthesized from the last known fiducial measurement projected forward by the calibration pixel. A gain history fits file meeting the requirements of the \rsl energy assignment {\tt rslpha2pi} tool was constructed using the time- and pixel-dependent effective temperature. The archived data was then reprocessed using the {\tt xapipeline} reprocessing script with the {\it calmethod} parameter set to `Cal-pix' and the {\it userghf} parameter set to the custom gain history file. In this mode, the script will construct a gain history file in the usual way (which is why the calibration pixel method was chosen), but will apply the file designated by the {\it userghf} setting.

\section{XSPEC Local Models}

We develop {\tt Xspec} convolution models that empirically reproduce asymmetrically broadened or Compton down-scattered 
spectra\footnote{These codes are available on https://github.com/HEASARC/xspec\_localmodels/tree/master/redistmodels}.
These models utilize the calcManyLines generic code in {\tt Xspec}, used in the {\tt gsmooth} function and other smoothing functions.

\subsection{Htsmooth}
\label{app:htsmooth}

This model, {\tt htsmooth} (Half-Triangle SMOOTH, Figure~\ref{fig:htsmooth} {\it left}),
redistributes the flux within each spectral bin (Ebin) to a half-triangular distribution.
The model includes a single parameter, MaxVel, 
in the unit of \UNITVEL, which represents the maximum radial velocity in the absolute
value at the end of the half-triangle. 
A negative MaxVel value indicates a blueshift, 
while a positive value indicates a redshift, 
extending the triangle into the higher or lower energy range, 
respectively.

\subsection{Emcomp}
\label{app:emcomp}

\begin{figure}
\plotone{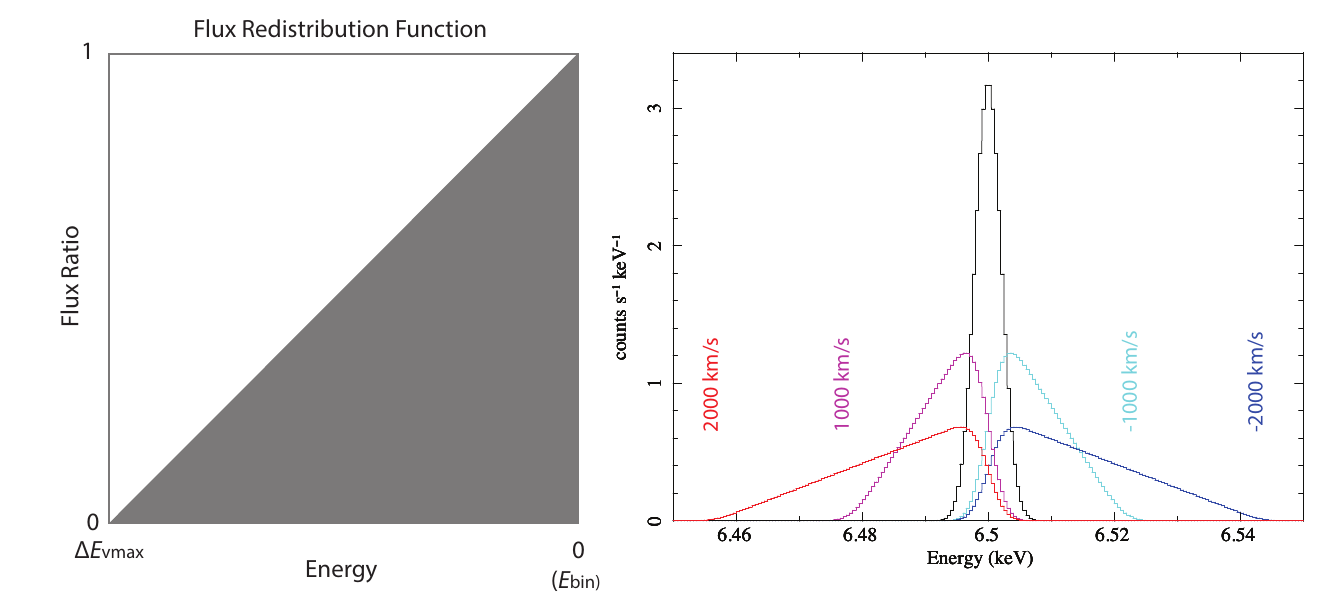}
\caption{
{\it Left}: Redistribution function of the {\tt htsmooth} model.
The corresponding energy of the free model parameter, $v_{\rm max}$,
is indicated on the figure.
{\it Right}: The function applied to a Gaussian line at 6.5~keV ({\it black}).
The MaxVel parameter for each model is displayed in its designated color.
All models are convolved with a response function for \rsl Hp events.
\label{fig:htsmooth}
}
\end{figure}

This model, {\tt emcomp} (EMpirical COMPton, Figure~\ref{fig:htsmooth} {\it right}),
redistributes the flux within each spectral bin (Ebin) to 
a trapezoidal distribution, defined down to the 180\DEGREE\ Compton backscattering energy ($E_{\rm BS}$) using the Smallest Scattering Fraction (SSF), Largest Scattering Fraction (LSF), and Smallest Scattering Cut (SSC). The 180-degree Compton backscattering energy $E_{\rm BS}$ is:

\begin{equation}\label{eq:BS}
E_{\rm BS} = \frac{E_{\rm bin}}{1+2(E_{\rm bin}/511~{\rm keV})}     
\end{equation}

\noindent The Compton Flux Ratio (CFR) parameter is the ratio of the Compton-scattering flux to the source flux.
After conducting preliminary trials, the LSF parameter is fixed at 1 to prevent any 
issues with degeneracy.
This model is empirical and does not account for the angular dependence of the Compton scattering cross-section. 
It is important to note that in a cold medium, photoelectric absorption occurs with Compton scattering. 
The Compton scattering cross-section is relatively constant within the \XRISM\ X-ray energy range, while the absorption cross-section is smaller at a higher energy. 
Consequently, the optical depth changes, and the total Compton-scattering flux changes with energy. 
This model does not account for this effect and, therefore, is not suitable for wide-band fitting.

    \begin{figure}
    \plotone{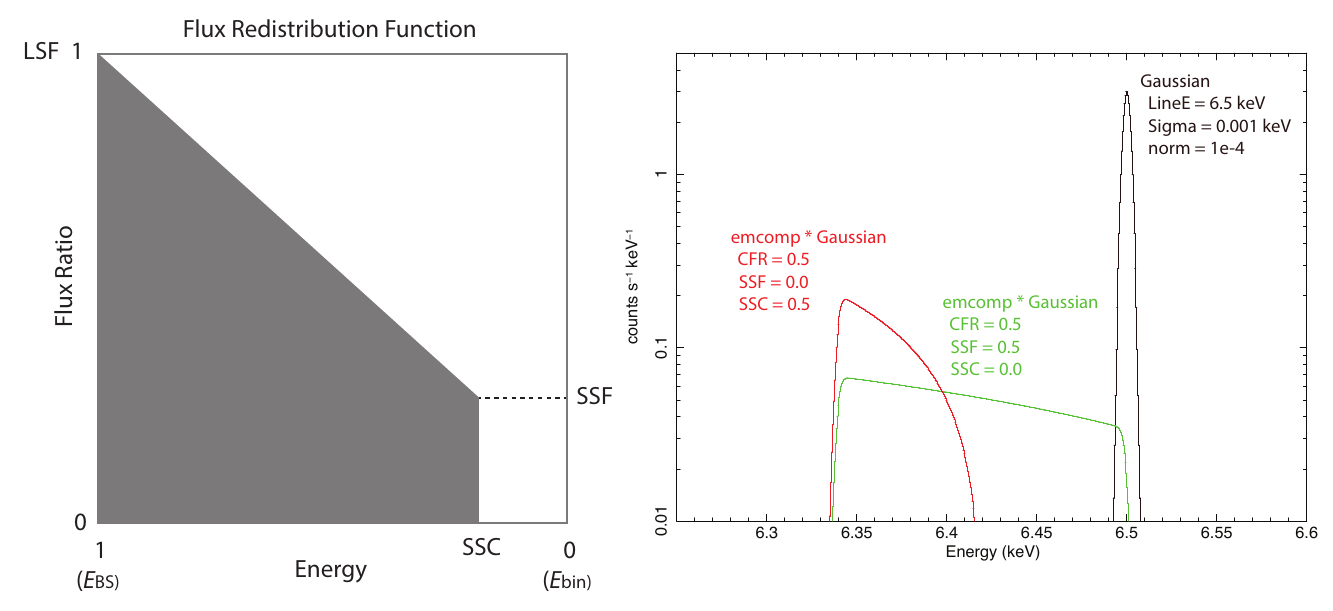}
    \caption{{\it Left}: Redistribution function of the {\tt emcomp} model.
    The free model parameters SSC, SSF, and LSF are indicated on the figure.
    {\it Right}: The function applied to a Gaussian function at 6.5~keV ({\it black}).
    All models are convolved with a response function for \rsl Hp events.
    \label{fig:emcomp}
    }
    \end{figure}
    
\section{Best-Fit Non-X-ray Background Model}
\label{app:nxb}

The following table shows the best-fit parameters of the NXB v1 model (rsl\_nxb\_model\_v1.mo) to the NXB spectra,
which are extracted from the provisional NXB database collected between Dec. 25, 2023 and May 24, 2024.

\input{tab3}

\clearpage


\input{ms.bbl}
\end{document}

%% file: authorlist.tex

\author[0000-0003-4721-034X]{Marc Audard}
\affiliation{Department of Astronomy, University of Geneva, Versoix CH-1290, Switzerland} 
\email{Marc.Audard@unige.ch}

\author{Hisamitsu Awaki} 
\affiliation{Department of Physics, Ehime University, Ehime 790-8577, Japan}
\email{awaki@astro.phys.sci.ehime-u.ac.jp}

\author[0000-0002-1118-8470]{Ralf Ballhausen}
\affiliation{Department of Astronomy, University of Maryland, College Park, MD 20742, USA}
\affiliation{NASA Goddard Space Flight Center, Greenbelt, MD 20771, USA}
\affiliation{Center for Research and Exploration in Space Science and Technology (CRESST II), Greenbelt, MD 20771, USA}
\email{ballhaus@umd.edu}

\author[0000-0003-0890-4920]{Aya Bamba}
\affiliation{Department of Physics, University of Tokyo, Tokyo 113-0033, Japan}
\email{bamba@phys.s.u-tokyo.ac.jp}

\author[0000-0001-9735-4873]{Ehud Behar}
\affiliation{Department of Physics, Technion, Technion City, Haifa 3200003, Israel}
\email{behar@physics.technion.ac.il}

\author[0000-0003-2704-599X]{Rozenn Boissay-Malaquin}
\affiliation{Center for Space Sciences and Technology, University of Maryland, Baltimore County, Baltimore, MD, 21250 USA}
\affiliation{NASA Goddard Space Flight Center, Greenbelt, MD 20771, USA}
\affiliation{Center for Research and Exploration in Space Science and Technology (CRESST II), Greenbelt, MD 20771, USA}
\email{rozennbm@umbc.edu}

\author[0000-0003-2663-1954]{Laura Brenneman}
\affiliation{Center for Astrophysics | Harvard-Smithsonian, Cambridge, MA 02138, USA}
\email{lbrenneman@cfa.harvard.edu}

\author[0000-0001-6338-9445]{Gregory V.\ Brown}
\affiliation{Lawrence Livermore National Laboratory, Livermore, CA 94550, USA}
\email{brown86@llnl.gov}

\author[0000-0002-7762-3172]{Michael F.\ Corcoran}
\affiliation{The Catholic University of America, 620 Michigan Avenue NE, Washington, DC 20064, USA}
\affiliation{NASA Goddard Space Flight Center, Greenbelt, MD 20771, USA}
\affiliation{Center for Research and Exploration in Space Science and Technology (CRESST II), Greenbelt, MD 20771, USA}
\email{corcoranm@cua.edu}

\author[0000-0002-5466-3817]{Lia Corrales}
\affiliation{Department of Astronomy, University of Michigan, Ann Arbor, MI 48109, USA}
\email{liac@umich.edu}

% ...


%% file: tab1.tex
\begin{deluxetable}{lccccccc}
\tablecolumns{8}
\tablewidth{0pc}
\tablecaption{Log of the \XRISM\ \rsl\ Observations\label{tbl:obslogs}}
\tablehead{
\colhead{Obs.}&
\colhead{Obs. ID}&
\multicolumn{2}{c}{Start/Stop Time}&
\colhead{Duration}&
\colhead{Exposure}&
\colhead{CR}&
\colhead{$v_{\earth}$}\\ \cline{3-4}
&&\colhead{Time}&\colhead{Phase}&\colhead{(ksec)}&\colhead{(ksec)}&\colhead{(\UNITCPS)}&\colhead{(\UNITVEL)}
}
\startdata
231123&000121000&2023 Nov.~23 02:06 ... 24 03:00&4.6811...6&89.7&44.2&0.76&9.8\\
240609&300067010&2024 June~9 01:17 ... 13 07:41&4.7794...815&368.7&296.0&1.1&$-$12.8\\
\enddata
\tablecomments{
Start/Stop Time: time interval with \rsl\ data after the standard screening.\\
Phase: binary orbital phase derived from the ephemeris in \citet{Corcoran2017a} equation 4.\\
CR: \rsl\ Hp count rate between 1.7$-$12.0~keV.\\
$v_{\earth}$: barycentric radial velocity of the Earth to \etacar,
calculated with the \texttt{radial\_velocity\_correction} method in 
the \texttt{SkyCoord} class of the \texttt{astropy coordinates} package. 
The positive sign is when the Earth is moving toward the star.
}
\end{deluxetable}

%% file: tab2.tex
\begin{deluxetable}{lccc}
\tablecolumns{4}
\tablewidth{0pc}
\tablecaption{Best-Fit Result\label{tbl:bestfitspec}}
\tablehead{
\colhead{Parameters}&
\colhead{Unit}&
\colhead{XRI231123}&
\colhead{XRI240609}
}
\startdata
Thermal Emission\\
~~\KT 			& (keV) 					& 4.01 (3.79...4.25) 			& 4.060 (4.059...4.062) \\
~~Abundance 	& (solar) 						& 0.68 (0.62...0.79) 			& 0.71 (0.69...0.73)\\
~~Redshift$^{\dagger}$ 		& (km/s)				& 12.8$^{\ast}$ ($-$85.6...81.6)		& $-$12.8 ($\pm$27$^{\ddagger}$)\\
~~Narrow\\
~~~~Broadening ($v_{\sigma}^{\rm N}$)	& (km/s)			& 332$^{\ast}$ (269...467) 				& 293 (284...301)\\
~~~~Normalization	& (10$^{-2}$) 				& 2.32 (1.64...3.63) 			& 3.69 (3.56...3.85)\\
~~Blue Wing\\
~~~~Maximum Velocity$^{\dagger}$ & (km/s)				& $-$2142 ($-$2387...$-$2009) 	& $-$2220 ($-$2305...$-$2136)\\
~~~~Normalization	& (10$^{-2}$) 				& 2.90 (2.16...3.18) 			& 2.97 (2.89...3.04)\\
~~Red Wing\\
~~~~Maximum Velocity$^{\dagger}$ & (km/s)				& 2922 (2550...4482) 			& 3107 (2985...3237)\\
~~~~Normalization	& (10$^{-2}$) 				& 1.93 (1.12...2.52) 			& 3.85 (3.77...3.94)\\
Fe Fluorescence\\
~~Broadening ($v_{\sigma6}$)	& (km/s)					& 319$^{\ast}$ (271...417) 				& 181 (172...188)\\
~~Redshift$^{\dagger}$ 	& (km/s) 					& $-$65$^{\ast}$ ($-$129...23.4) 			& $-$126 ($-$135...$-$118)\\
~~K$\alpha_1$ flux & (10$^{-5}$ ph/cm$^{2}$/s)	& 2.21 (1.90...2.63) 			& 2.93 (2.79...3.08)\\
~~K$\alpha_2$ flux & (10$^{-5}$ ph/cm$^{2}$/s) 	& 1.48 (=1.31~K$\alpha_1$)	& 1.96 (1.91...2.02)\\
~~K$\beta$ flux & (10$^{-5}$ ph/cm$^{2}$/s) 		& 0.49 (0.31...0.89) 			& 0.44 (0.42...0.45)\\
\multicolumn{3}{l}{Compton Scattering}\\
~~CFR & 								& 0.103 (0.081...0.136) 			& 0.080 (0.074...0.086)\\
~~SSF & (10$^{-4}$)							& 6.2 (fix) 				& 6.2 (5.5...6.9)\\
~~SSC & 								& 0.38 (fix)		 			& 0.38 (0.36...0.40)\\ \hline
C-Statistic (dof) & 							& 2155.23 (1786) 				& 1792.64 (1783)\\
\enddata
\tablecomments{
The parentheses show 90\% confidence ranges.\\
$^{\dagger}$These radial velocities are corrected for the solar barycentric system.\\
$^{\ddagger}${\tt Xspec/chain} derives an unreasonably small confidence range,
perhaps due to an erroneous fitting behavior near zero.
This error range is for the redshift parameter of a 1$T$ {\tt apec} fit 
fixing the broadening parameter to the 4-band spectrum, described in Section~\ref{subsec:iron_Ka_lines}.\\
$^{\ast}$These values have additional systematic uncertainties at $\lesssim$1~eV ($\sim$45~\UNITVEL).\\
CFR: Compton Flux Ratio, SSF: Smallest Scattering Fraction, SSC: Smallest Scattering Cut.\\
}
\end{deluxetable}

%% file: acknowledgment.tex
\begin{acknowledgments}
We appreciate fruitful discussions with Drs.\ 
Augusto Damineli, Theodor R.~Gull, D.~John Hillier, Jonathan Mackey, Anthony F.J., Moffat, Ya{\"e}l Naz{\'e}, 
Julian M.~Pittard, Noel Richardson, Christopher Russell, and Gerd Weigelt.
The material is based upon work supported by NASA under award numbers 
80GSFC21M0002, 80GSFC24M0006,
80NSSC18K1684, 80NSSC18K0988, 
80NSSC20K0733, 80NSSC20K0737, 80NSSC20K0883, 
80NNSC22K1922,
80NSSC23K1656, 
80NSSC24K0678, 80NSSC24K1148, 80NSSC24K1774,
80NSSC25K7064, and 80NSSC25K7536,
and contract NAS8-0360.
This work is supported by JSPS KAKENHI grant numbers 
JP19K14762, JP19K21884, 
JP20H00157, JP20H01946, JP20H01947, JP20H05857, JP20K04009, JP20K14491, JP20KK0071, 
JP21H01095, JP21H04493, JP21K03615, JP21K13958, JP21K13963, 
JP22H00158, 
JP23H00121, JP23H00151, JP23H01211, JP23K03454, JP23K03459, JP23H04899, JP23K13154, JP23K20239, JP23K20850, JP23K22548, JP23KJ1807,
JP24H00253, JP24K00638, JP24K00672, JP24K00677, JP24K17093, JP24K17104, JP24K17105, 
JP25H00672,  and JP25K01026.
This work is supported by
Lawrence Livermore National Laboratory under Contract DE-AC52-07NA27344,
STFC through grant ST/T000244/1,
Canadian Space Agency grant 18XARMSTMA,
the Alfred P. Sloan Foundation through the Sloan Research Fellowship,
the JSPS Core-to-Core Program, JPJSCCA20220002,
the RIKEN Pioneering Project Evolution of Matter in the Universe (r-EMU),
the RIKEN SPDR Program,
Rikkyo University Special Fund for Research (Rikkyo SFR),
the Kagoshima University postdoctoral research program (KU-DREAM),
the Strategic Research Center of Saitama University,
Program for Forming Japan's Peak Research Universities (J-PEAKS), and
the Organization for the Promotion of Gender Equality at Nara Women’s University.
\end{acknowledgments}

%% file: tab3.tex
\begin{deluxetable}{lcc}
\tablecolumns{3}
\tablewidth{0pc}
\tablecaption{NXB Model\label{tbl:bestfitnxb}}
\tablehead{
\colhead{Parameter}&
\colhead{XRI231123}&
\colhead{XRI240609}
}
\startdata
1 & 		0.990114 		&	0.965222\\
3 &		8.21591E-04 	&	8.21591E-04\\
7 &		3.84967E-05 	&	4.16577E-05\\
14 & 		2.86589E-05 	&	2.11130E-05\\
20 &		4.82967E-05	&	3.72466E-05\\
23 &		1.11295E-05	&	1.24417E-05\\
29 &		8.19269E-06	&	1.22293E-05\\
35 &		3.97141E-05	&	4.50395E-05\\
41 &		1.75562E-05	&	2.11823E-05\\
47 &		2.22964E-05	&	1.43713E-05\\
50 &		1.58155E-04	&	1.70030E-04\\
53 &		1.40780E-04	&	1.42236E-04\\
56 &		4.16170E-05	&	4.85166E-05\\ \hline
\enddata
\tablecomments{
Par: the sequential parameter numbers of the NXB v1 model, which appear in {\tt Xspec} outputs.
}
\end{deluxetable}